\documentclass[sigconf]{acmart}
\AtBeginDocument{%
  \providecommand\BibTeX{{%
    \normalfont B\kern-0.5em{\scshape i\kern-0.25em b}\kern-0.8em\TeX}}}

\usepackage{paracol}
\globalcounter{algocf}
\usepackage[bottom]{footmisc}
\usepackage{array}
\usepackage{wrapfig}
\usepackage{stmaryrd}

\newcolumntype{P}[1]{>{\centering\arraybackslash}p{#1}}
\usepackage{adjustbox}
\usepackage{graphicx}
\usepackage{algorithmic}
\usepackage{textcomp}
\usepackage{comment}
\usepackage{tikz}
\usetikzlibrary{automata, positioning, arrows, shapes.misc, patterns}
\usepackage{multirow}
\usepackage{enumitem}
\usepackage{graphics}
\usepackage[ruled,vlined]{algorithm2e}
\usepackage{url}
\usepackage{hyperref}
\usepackage{amsmath}
\usepackage{amsmath}
\usepackage{hyperref}
\usepackage[english]{babel}
\usepackage{caption}
\usepackage{subcaption}
\usepackage{multirow}
\usepackage{multicol}
\usepackage{cleveref}
\usepackage{graphicx}
\usepackage{mathtools}
\usepackage{lineno}
\usepackage{siunitx}
\usepackage{pifont}

\usepackage{ragged2e}
\newlength\myindent
\graphicspath{{figs/}}

\newcommand{\states}{\mathrm{\Sigma}}
\newcommand{\statespace}{\states}

\newcommand{\Finally}{\mathbf{F}}
\newcommand{\Globally}{\mathbf{G}}

\newcommand{\AP}{\mathit{AP}}
\newcommand{\trace}{t}

\newcommand{\E}{\mathbf{E}}
\newcommand{\A}{\mathbf{A}}
\newcommand{\zplus}{\mathbb{Z}_{\geq 0}}

\newcommand{\DD}{{d}}
\newcommand{\TT}{\mathbb{T}}

\newcommand{\HH}{\mathbf{H}}

\newcommand{\ser}{\odot}

\definecolor{bln_blue}{HTML}{00A8FF}
\definecolor{bln_red}{HTML}{c23616}
\definecolor{bln_green}{HTML}{16A085}
\definecolor{bln_magenta}{HTML}{9B59B6}
\definecolor{blond}{rgb}{0.98, 0.94, 0.75}
\definecolor{beige}{rgb}{0.96, 0.96, 0.86}
\definecolor{babyblueeyes}{rgb}{0.63, 0.79, 0.95}
	\definecolor{beaublue}{rgb}{0.74, 0.83, 0.9}
 
\newcommand{\LN}[1]{{\textcolor{blue}{LN: #1}}}

 \usepackage{algorithm2e}
\SetAlFnt{\normalsize}
\SetAlCapFnt{\normalsize}
\SetAlCapNameFnt{\normalsize}
\algsetup{linenosize=\normalsize}

\copyrightyear{2023} 
\acmYear{2023} 
\setcopyright{acmlicensed}\acmConference[MEMOCODE '23]{21st ACM-IEEE International Conference on Formal Methods and Models for System Design}{September 21--22, 2023}{Hamburg, Germany}
\acmBooktitle{21st ACM-IEEE International Conference on Formal Methods and Models for System Design (MEMOCODE '23), September 21--22, 2023, Hamburg, Germany}
\acmPrice{15.00}
\acmDOI{10.1145/3610579.3611077}
\acmISBN{979-8-4007-0318-8/23/09}

\begin{document}

\title{Model Checking Time Window Temporal Logic for Hyperproperties}

\author{Ernest Bonnah}

\affiliation{%
  \institution{University of Missouri-Columbia}
  \streetaddress{411 S 6th St}
  \city{Columbia}
  \state{Missouri}
  \country{USA}
  \postcode{65211}}
 \email{ernest.bonnah@mail.missouri.edu}

\author{Luan Viet Nguyen}
\affiliation{%
  \institution{University of Dayton}
  \city{Dayton}
  \state{Ohio}
  \country{USA}}
\email{lnguyen1@udayton.edu}

\author{Khaza Anuarul Hoque}
\affiliation{%
 \institution{University of Missouri-Columbia}
 \streetaddress{411 S 6th St}
 \city{Columbia}
 \state{Missouri}
 \country{USA}}
\email{hoquek@umsystem.edu}


\begin{abstract}
Hyperproperties extend trace properties to express properties of sets of traces, and they are increasingly popular in specifying various security and performance-related properties in domains such as cyber-physical systems, smart grids, and automotive. This paper introduces HyperTWTL, which extends Time Window Temporal Logic (TWTL)--a domain-specific formal specification language for robotics, by allowing explicit and simultaneous quantification over multiple execution traces. We propose two different semantics for HyperTWTL, \emph{synchronous} and \emph{asynchronous}, based on the alignment of the timestamps in the traces. Consequently, we demonstrate the application of HyperTWTL in formalizing important information-flow security policies and concurrency for robotics applications. Furthermore, we introduce a model checking algorithm for verifying fragments of HyperTWTL by reducing the problem to a TWTL model checking problem.
\end{abstract}




\maketitle

\section{Introduction}
\label{intro}
Hyperproperties \cite{clarkson2010hyperproperties} extend the notion of trace properties  \cite{alpern1985defining} from a set of traces to a set of sets of traces. In other words, a hyperproperty specifies system-wide properties in contrast to the property of just individual traces. This allows us to specify a wide range of properties related to information-flow security~\cite{zdancewic2003observational,goguen1982security}, consistency models in concurrent computing~\cite{bonakdarpour2020controller,finkbeiner2017eahyper}, robustness models in cyber-physical systems~\cite{bonakdarpour2018monitoring,garey1979computers}, and also service level agreements (SLA)~\cite{clarkson2010hyperproperties}. Motivated by the expressiveness of hyperproperties, several hyper-temporal logics, such as HyperLTL~\cite{clarkson2014temporal}, HyperSTL~\cite{nguyen2017hyperproperties}, and HyperMTL~\cite{bonakdarpour2020model}, were recently proposed by extending the conventional temporal logics such as Linear Temporal Logic (LTL)~\cite{pnueli1977temporal}, Signal Temporal Logic (STL)~\cite{maler2004monitoring}, and Metric Temporal Logic (MTL)~\cite{koymans1990specifying}, respectively. Consequently, various model checking techniques have been proposed for verifying HyperLTL \cite{clarkson2014temporal, coenen2019verifying, finkbeiner2018model, finkbeiner2015algorithms}, HyperMTL \cite{bonakdarpour2020model,ho2018verifying}, HyperMITL \cite{ho2021timed}, HyperSTL \cite{nguyen2017hyperproperties} specifications employing alternating automata, model checking, strategy synthesis, and several other methods \cite{hsu2021bounded}.

\begin{table}[b!]
\centering
\caption{The representation of $\varphi$ in HyperTWTL, HyperSTL, HyperMTL}\label{ex_reqs} 
\setlength{\tabcolsep}{1pt}
\begin{tabular}{|ll|}
\hline
HyperTWTL&$\forall \pi \forall \pi' \cdot [\HH^{5}~A_{\pi}]^{[0,6]} \ser [\HH^{2}~B_{\pi'}]^{[7,10]}$\\
HyperSTL&$ \forall \pi \forall \pi' \cdot ({\Finally}_{[0,6-5]}{\Globally}_{[0,5]} A_{\pi}) \land ({\Finally}_{[7,10-2]} {\Globally}_{[0,2]} B_{\pi'})$\\
HyperMTL&$ \forall \pi \forall \pi' \cdot {\bigvee}_{i=0}^{6-5}({\Globally}_{[i,i+5]} A_{\pi} \land  {\bigvee}_{j=i+5+7}^{i+5+10-2} {\Globally}_{[j,j+2]} B_{\pi'})$\\
\hline
\end{tabular}
\end{table}
Time bounded specifications are common in many applications, such as robotics. Such specifications with time constraints cannot be expressed using LTL, however they can be expressed using Metric Temporal Logic (MTL) \cite{koymans1990specifying}, Signal Temporal Logic (STL)\cite{maler2004monitoring}, Bounded Linear Temporal Logic (BLTL)\cite{tkachev2013formula} and Time Window Temporal Logic (TWTL)~\cite{vasile2017time}. The TWTL has rich semantics and is known for expressing specification more compactly when compared to Metric Temporal Logic (MTL) \cite{koymans1990specifying}, Signal Temporal Logic (STL)\cite{maler2004monitoring}, Bounded Linear Temporal Logic (BLTL)\cite{tkachev2013formula}. For instance, let us consider a specification as ``\textit{stay at Q for the first 5-time steps within the time window [0, 10]}”. This can be expressed in TWTL as \texttt{$[{\HH}^5 Q]^{[0,10]}$}. The exact specification can be expressed in STL as \texttt{${\Finally}_{[0,10-5]}{\Globally}_{[0,5]} Q$} where the outermost time window needs to be modified with respect to the inner time window. Furthermore, TWTL presents an explicit concatenation operator ($\ser$), which is very useful in expressing serial tasks in robotic mission specification and planning. Thus, TWTL is becoming popular in different application domains \cite{bonnah2022runtime, asarkaya2021temporal, peterson2021distributed, aksaray2021probabilistically, buyukkoccak2021distributed, asarkaya2021persistent}.  However, the conventional TWTL can only express trace properties (can reason about individual traces). This limits their application to evaluating a wide range of security properties as evaluating them requires reasoning about multiple traces. A straightforward extension of TWTL to a hyper-temporal logic will result in a flawed formalism and impractical for robotic applications. There are multiple challenges in designing a domain-specific temporal logic for timed hyperproperties. First, the alignment of time stamps across multiple traces of a system must be taken into consideration. This is due to the complexity of interpreting time stamps that are not aligned across traces. Additionally, the framework of any hyper-temporal logic has to consider the speed of time with which time-stamped traces proceed. This is because traces may not proceed at the same speed of time. To reason about traces that move at different speeds of time, \emph{asynchronous semantics} need to be allowed.

A hyper-temporal logic, HyperTWTL, extends the classical semantics of TWTL with trace variables and explicit existential and universal quantifiers over multiple execution traces. HyperTWTL can express bounded hyperproperties more compactly when compared to HyperMTL and HyperSTL. For instance, consider a hyperproperty that requires that ``\emph{for any pair of traces $\pi$ and $\pi$' of a system, $A$ should hold for 5-time steps in trace $\pi$ within the time bound $[0, 6]$, afterwards B should also hold for 2-time steps in trace $\pi$' within the time bound $[7, 10]$}.'' This requirement $\varphi$ can be expressed using HyperTWTL, HyperSTL, and HyperMTL formalisms as shown in Table~\ref{ex_reqs}. The compact syntax of HyperTWTL allows for a more succinct representation of this requirement than HyperMTL and HyperSTL, which require nested operators, shifted time windows, and the disjunction of several sub-formulae. 
For instance, in Table \ref{ex_reqs}, it can also be observed that the given requirement can be formalized in HyperTWTL formula with total 5 temporal operators (without considering the quantifiers). This same requirement can be formalized as a HyperMTL formula using 11 temporal operators (excluding the quantifiers). This example shows the succinct characteristics of HyperTWTL over HyperMTL. Indeed, with the increasing complexity of requirements, the complexity of HyperMTL formulae will also grow, which makes the formal analysis of HyperMTL formulae very expensive for complicated robotic applications.


In this paper, we describe a full version of HyperTWTL with two semantics, \emph{synchronous} and \emph{asynchronous}, to represent system behaviors as a sequence of events that can occur either at the same or different time points respectively. In the synchronous semantics, the operators are evaluated time-point by time-point similar to HyperLTL. In contrast, asynchronous semantics allows for evaluation over traces that proceed at different speed of time.  We demonstrate how HyperTWTL can be used to formalize important security policies such as non-interference, opacity, countermeasures to side-channel timing attacks, and also concurrency-related properties, such as linearizability. Such security requirement are common not only in robotics but also in other computing domains at hardware, software and system level.
Finally, we propose a model checking algorithm for fragments (alternation-free and $k$-alternations) of HyperTWTL by reducing the model checking problem to the TWTL model checking problem.

The rest of the paper is organized as follows: Section~\ref{prelims} reviews the preliminary concepts. The syntax and semantics of HyperTWTL are presented in Section~\ref{HyperTWTL}. In Section~\ref{specs}, some applications of HyperTWTL are discussed. 
Our model checking algorithm for HyperTWTL is presented in section~\ref{model checking}. We evaluate the feasibility of our proposed algorithm in Section~\ref{specs}. Related works are discussed in Section~\ref{related-works}. Finally, Section~\ref{conclusion} concludes the paper.

\section{Preliminaries}
\label{prelims}
Let $\AP$ be a finite set of {\em atomic propositions} and $\statespace
= 2^{\AP}$ be the powerset of $\AP$. Let $\mathcal{A} = \zplus \times \Sigma$ be the {\em alphabet}, where $\zplus$ is the set of non-negative integers. A {(time-stamped) \em event} is a member of the alphabet $\mathcal{A}$ and is of the form $(\tau, e)$, where $\tau \in \zplus$ and $e \in \statespace$. A trace $\trace \in \mathcal{A}^{\omega}$ denotes an infinite sequence of events over $\mathcal{A}$, and $\trace \in \mathcal{A}^{*}$ denotes a finite sequence of events over $\mathcal{A}$. For a trace $t$, we denote $\trace[n].e$ as the event at time $n$, i.e. $e_n$ and by $\trace[n].\tau$, we mean $\tau_n$. Let $t[i, j]$ denote the subtrace of trace $\trace$ starting from time $i$ up to time $j$. We model timed systems as Timed Kripke Structures with the assigned time elapsed on the transitions. \\

\noindent \textbf{Definition 1.} A timed Kripke structure (TKS) is a tuple $\mathcal{T} = (S, S_{init}, \delta, AP, L)$ where  
\begin{itemize}
    \item $S$ is a finite set of states;
    \item $S_{init} \subseteq S$ is the set of initial states;
    \item $\delta \subseteq S \times \zplus \times S$ is a set of transitions;
    \item $AP$ is a finite set of atomic propositions; and 
    \item $L:S \rightarrow \Sigma$ is a labelling function on the states of $\mathcal{T}$
\end{itemize} 
   
We require that for each $s \in S$, there exists a successor that can be reached in a finite number of transitions. Hence, all nodes without any outgoing transitions are equipped with self-loops such that $(s, 1, s) \in \delta$. A path over an TKS is an infinite sequence of states $s_{0}s_{1}s_{2}\dots \in S^{\omega}$, where $s_0 \in S_{init}$ and $(s_{i}, d_{i}, S_{i+1})\in\delta$, for each $i \geq 0$. A  trace  over TKS is of the form: $\trace =  (\tau_0, e_0) (\tau_1, e_1) (\tau_2, e_2) \dots$, such that there exists a path $s_{0}d_{0}s_{1}d_{1}s_{2}d_{2}\cdots \in S^{\omega}$. \\

\noindent \textit{\textbf{TWTL Syntax and Semantics.}} 
The set of TWTL formulae over a finite set of atomic propositions is defined by the following syntax: 
\begin{equation*}
\begin{aligned}
& \phi:= \top \mid \HH^{\DD} a \mid  \HH^{\DD} \neg a \mid \phi_1 \wedge \phi_2 \mid \neg \phi \mid \phi_1 \ser \phi_2 \mid [\phi]^{[\tau, \tau']}
    \end{aligned}
\end{equation*}
\noindent where $\top$ stands for true, $a$ is an atomic proposition in $AP$. The operators $\HH^{\DD}$, $\ser$ and $[\  ]^{[\tau,\tau']}$ represent the hold operator with $\DD \in \zplus$, concatenation operator and within operator respectively within a discrete-time constant interval $[\tau, \tau']$, where $\tau, \tau' \in \zplus$ and $\tau' \geq \tau$, respectively and $\wedge$ and $\neg$ are the conjunction and negation operators respectively. The disjunction operator ($\vee$) can be derived from the negation and conjunction operators. Likewise, the implication operator ($\rightarrow$) can also be derived from the negation and disjunction operators.\\ \\
\noindent \textit{TWTL semantics}: The satisfaction relation defined by $\models$ defines when subtrace $\trace[i,j]$ of a (possibly infinite) timed-trace $\trace$ from position $i$ up to and including position $j$, satisfies the TWTL formula. This is denoted by $\trace[i,j] \models \phi$. The formula $\phi = \top$ always holds. The hold operator $\phi = \HH^{\DD} a$ requires that  $a$ should hold for $\DD$ time units. Likewise $\HH^{\DD} \neg a$, specifies that for $\DD$ time units, the condition $a \notin \trace[n].e$ should hold. The trace $\trace[i,j]$ satisfies the formulae $\phi = \phi_1 \land \phi_2$  when both subformulae are satisfied while in  $\neg \phi$, $\trace[i,j]$, does not satisfy the given formula. A given formula in the form $\phi_1 \ser\phi_2$ specifies that the $\trace[i,j]$ should satisfy the first formula first and the second afterward with one time unit difference between the end of execution of $\phi_1$ and start of execution of $\phi_2$. The trace $\trace[i,j]$,  must satisfy $\phi$ between the time window $[\tau, \tau']$ given $[\phi]^{[\tau, \tau']}$. 

Given a TWTL formula $\phi$ and a timed-trace, $\trace[i,j]$ where the trace starts at $\tau_i \geq 0$ and terminates at $\tau_j \geq \tau_i$, the semantics of the operators is defined as:
\[
\begin{array}{l@{\hspace{1em}}c@{\hspace{1em}}l}
\trace[i,j] \models \top \\
\trace[i,j] \models \HH^{\DD} a  & \text{iff} & \text{$a \in t[n].e$, $\forall n \in \{i,...,i + \DD\} \wedge$} \\ 
&  & \text{$(t[j].\tau - t[i].\tau) \geq \DD$} \\
\trace[i,j] \models \HH^{\DD} \neg a  & \text{iff} & \text{$a \notin t[n].e$, $\forall n \in \{i,...,i + \DD\} \wedge $} \\
& & \text{$(t[j].\tau - t[i].\tau) \geq \DD$} \\
\trace[i,j] \models  \phi_1 \wedge \phi_2  & \text{iff} & \text{$(\trace[i,j] \models \phi_1) \wedge (\trace[i,j] \models \phi_2)$} \\
\trace[i,j] \models  \neg \phi  & \text{iff} & \text{$\neg (\trace[i,j] \models \phi)$} \\
\trace[i,j] \models \phi_1 \ser\phi_2 &  \text{iff} &\text{$\exists k = \arg\min_{i \le k < j}\{\trace[i,k] \models \phi_1\}\wedge$} \\
& & \text{$(\trace[k + 1,j] \models \phi_2)$} \\
\trace[i,j] \models [\phi]^{[\tau, \tau']}   & \text{iff} & \text{$\exists k \geq i + \tau, ~s.t.~ \trace[k,i + \tau'] \models \phi ~ \wedge$} \\ & & \text{$(t[j].\tau - t[i].\tau) \geq \tau'$} 
\end{array}
\]

\section{HyperTWTL}
\label{HyperTWTL}
A {\em hyperproperty} is a set of sets of infinite traces. HyperTWTL is a hyper-temporal logic to specify hyperproperties for TWTL~\cite{vasile2017time} by extending the TWTL with quantification over multiple and concurrent execution traces. Hyperproperties can be specified with HyperTWTL using two different semantics, \textit{synchronous} and \textit{asynchronous}. Synchronous semantics requires timestamps in all quantified traces to match at each point in time. However, we can reason about traces that proceed at different speeds with asynchronous semantics. We first present the syntax and then the synchronous and asynchronous semantics of HyperTWTL.

\subsection{Syntax of HyperTWTL}
The syntax of HyperTWTL is inductively defined by the grammar as follows.
\begin{equation*}
\begin{aligned}
& \varphi := \exists\pi \cdot\varphi \mid \forall\pi \cdot\varphi \mid \phi\\
& \phi := \HH^{d} a_{\pi} \mid  \HH^{d} \neg a_{\pi} \mid \phi_1 \wedge \phi_2 \mid  \neg \phi \mid \phi_1 \ser \phi_2 \mid [\phi]^{[\tau, \tau']} \mid \\ & ~~~~~~ \E\rho \cdot\psi \mid \A\rho \cdot\psi \\
& \psi := \HH^{d} a_{\pi, \rho} \mid  \HH^{d} \neg a_{\pi, \rho} \mid \psi_1 \wedge \psi_2 \mid  \neg \psi \mid \psi_1 \ser \psi_2 \mid [\psi]^{S, T}
\end{aligned}
\vspace{-0.4mm}
\end{equation*}

where $a \in AP$, $\pi$ is a trace variable from a set of trace variables ${\mathcal{V}}$ and $\rho$ is a trajectory variable from the set $\mathcal{P}$. Thus, given $a_{\pi, \rho}$, the proposition $a \in AP$ holds in trace $\pi$ and trajectory $\rho$ at a given time point. The quantified formulae $\exists \pi$, and $\forall \pi$ are interpreted as ``there exists some trace $\pi$'' and ``for all the traces $\pi$'', respectively. The operators $\HH^{d}$, $\ser$ and $[\  ]^{[\tau, \tau']}$ represent the hold operator with $d \in \zplus$, concatenation operator and within operator respectively with a discrete-time constant interval $[\tau, \tau']$, where $\tau, \tau' \in \zplus$ and $\tau' \geq \tau$, respectively and $\wedge$ and $\neg$ are the conjunction and negation operators respectively. Trace quantifiers $\exists\pi$ and $\forall \pi$,  allow for the simultaneous reasoning about different traces. Similarly, trajectory quantifiers $\E \rho$ and $\A \rho$ allow reasoning simultaneously about different trajectories. The quantifier $\E$ is interpreted as there exists a trajectory that gives an interpretation of the relative passage of time between the traces for the inner temporal formula to be evaluated. Similarly, $\A$ means that all trajectories satisfy the inner formula. $S$ and $T$ are both intervals of form $[\tau, \tau']$ where $\tau, \tau' \in \zplus$ and $\tau' \geq \tau$. The disjunction operator ($\vee$) can be derived from the negation and conjunction operators. Likewise, the implication operator ($\rightarrow$) can also be derived from the negation and disjunction operators.

\begin{table*}[ht]
\centering
\caption{\label{Sync}\textcolor{black}{Synchronous semantics of HyperTWTL}}
\begin{tabular}{|lll|}
\hline
$(\TT, \Pi) \models_s \exists \pi. \varphi$ &  \text{iff} & $\exists \trace \in \TT \cdot (\TT, \Pi[\pi \rightarrow (t, 0)]) \models_s  \varphi$  \\
$(\TT, \Pi) \models_s \forall \pi. \varphi$ & \text{iff} &  $\forall \trace \in \TT \cdot (\TT, \Pi[\pi \rightarrow (t, 0)]) \models_s  \varphi$ \\
$(\TT, \Pi) \models_s \HH^d a_\pi $ &  \text{iff} &  \begin{tabular}[c]{@{}l@{}}\text{$a \in t[n].e$ for $(\trace,n) = \Pi(\pi)$, $\forall n \in$}  \text{$\{n,...,n + d\}$} \text{$\wedge ~(\trace[n+i].\tau$}  \text{$- ~\trace[n].\tau) \geq d$, for some $i >0$}\end{tabular}  \\
$(\TT, \Pi) \models_s \HH^d \neg a_\pi $ &  \text{iff} &  \begin{tabular}[c]{@{}l@{}}\text{$a \notin t[n].e$ for $(\trace,n) = \Pi(\pi)$, $\forall n \in$}  \text{$\{n,...,n + d\}$} \text{$\wedge ~(\trace[n+i].\tau$}  \text{$- ~\trace[n].\tau) \geq d$, for some $i >0$}\end{tabular}  \\
 $(\TT, \Pi) \models_s  \phi_1 \wedge \phi_2$ &  \text{iff} &  \text{$((\TT, \Pi)  \models_s \phi_1) ~\wedge ((\TT, \Pi)  \models_s \phi_2)$}  \\
 $(\TT, \Pi) \models_s  \neg \phi$ &  \text{iff} &  \text{$\neg ((\TT, \Pi) \models_s \phi)$}  \\
 $(\TT, \Pi) \models_s  \phi_1 \ser \phi_2$ & \text{iff} &  \begin{tabular}[c]{@{}l@{}}\text{$\exists k = \arg\min_{n \le k \le n'} \text{for some $i \in [n, k]$}: ((\TT, \Pi) \models_s \phi_1)$}, for~some $j \in [k+1, n']$: \text{$((\TT,  \Pi) \models_s \phi_2) $ }\end{tabular}  \\
$(\TT, \Pi) \models_s  [\phi]^{[\tau, \tau']}$ &  \text{iff} &  \begin{tabular}[c]{@{}l@{}}\text{$\exists k \geq n + \tau$}, s.t. for all {$i \in [k,n+\tau']$} \text{$(\TT, \Pi) \models_s \phi$} ~\text{$\wedge ((\Pi)+n' - \Pi) \geq \tau'$ for~some $n, n' \geq 0$}\end{tabular} \\ \hline
\end{tabular}
\end{table*}

\subsection{Semantics of HyperTWTL}
\label{semantics}
We classify the 
HyperTWTL formulae into two fragments based on the possible alternation in the HyperTWTL syntax as follows:
\begin{enumerate}
    \item Alternation-free HyperTWTL formulae with one type of quantifier, and
    \item $k$-alternation HyperTWTL formulae that allows $k$-alternation between existential and universal quantifiers. \\
\end{enumerate}

\subsubsection{Synchronous Semantics of HyperTWTL} The satisfaction relation gives the semantics of synchronous HyperTWTL $\models_s$ over time-stamped traces $\TT$. We define an assignment $\Pi:{\mathcal{V}} \rightarrow \mathcal{A}^{\omega} \times \zplus$ as a partial function mapping trace variables to time-stamped traces. Let $\Pi(\pi) = (\trace, n)$ denote the time-stamped event from trace $\trace$ at position $n$ currently employed in the consideration of trace $\pi$. We then denote the explicit mapping of the trace variable $\pi$ to a trace $\trace \in \TT$ at position $p$ as $\Pi[\pi \rightarrow (\trace, n)]$. Given $(\Pi )$, where $\Pi$ is the trace mapping, we use $(\Pi) + k$ as the $k^{th}$ successor of $(\Pi)$.  We now present the synchronous semantics of HyperTWTL in Table~\ref{Sync}.

In Table~\ref{Sync}, the hold operator $\HH^{d} a_{\pi}$ states that the proposition $a$ is to be repeated for $d$ time units in trace $\pi$. Similarly $\HH^{d} \neg a_{\pi}$, requires that for $d$ time units the proposition $a$ should not be repeated in trace $\pi$. The trace set $\TT$ satisfies both sub-formulae in $\phi = \phi_1 \wedge \phi_2$ while in  $\neg \phi$, $\TT$, does not satisfy the given formula. A given formula with a concatenation operator in the form $\phi_1 \ser\phi_2$ specifies that every $\trace \in \TT$ should satisfy $\phi_1$ first and then immediately $\phi_2$ must also be satisfied with one-time unit difference between the end of execution of $\phi_1$ and the start of execution of $\phi_2$. The trace set, $\TT$ must satisfy $\phi$ between the time window  within the time window $[\tau, \tau']$ given $[\phi]^{[\tau, \tau']}$.

\begin{table*}[ht]
\centering
\caption{\label{Async}\textcolor{black}{Asynchronous semantics of HyperTWTL}}
\begin{tabular}{|lll|}
\hline
$(\TT, \Pi, \Gamma) \models_a \exists \pi. \varphi$ &  \text{iff} & $\exists \trace \in \TT \cdot (\TT, \Pi[\pi, \rho \rightarrow t, 0], \Gamma) \models_a  \varphi$ for all $\rho$  \\
$(\TT, \Pi, \Gamma) \models_a \forall \pi. \varphi$ &  \text{iff} &  $\forall \trace \in \TT \cdot (\TT, \Pi[\pi, \rho \rightarrow t, 0], \Gamma) \models_a  \varphi$ for all $\rho$\\
$(\TT, \Pi, \Gamma) \models_a \E \rho. \varphi$ &  \text{iff} & $\exists v \in \mathcal{R}_{range(\Gamma)}: (\TT, \Pi, \Gamma[\rho \rightarrow v]) \models_a  \varphi$  \\
$(\TT, \Pi, \Gamma) \models_a \A \rho. \varphi$ &  \text{iff} &  $\forall v \in \mathcal{R}_{range(\Gamma)}: (\TT, \Pi, \Gamma[\rho \rightarrow v]) \models_a  \varphi$ \\
$(\TT, \Pi, \Gamma) \models_a \HH^d a_{\pi, \rho}$ & \text{iff} & \begin{tabular}[c]{@{}l@{}}\text{$a \in t[n].e$ for $(\trace,n) = \Pi(\pi, \rho)$, $\forall n \in$}  \text{$\{n,...,n + d\}$} \text{$\wedge ~(\trace[n+i].\tau$}  \text{$- ~\trace[n].\tau) \geq d$, for some $i >0$}\end{tabular} \\
$(\TT, \Pi, \Gamma) \models_a \HH^d \neg a_{\pi, \rho} $ & \text{iff} & \begin{tabular}[c]{@{}l@{}}\text{$a \notin t[n].e$ for $(\trace,n) = \Pi(\pi, \rho)$, $\forall n \in$}  \text{$\{n,...,n + d\}$} \text{$\wedge ~(\trace[n+i].\tau$}  \text{$- ~\trace[n].\tau)  \geq d$, for some $i >0$}\end{tabular}  \\
$(\TT, \Pi, \Gamma) \models_a  \psi_1 \wedge \psi_2$ & \text{iff} &  \text{$((\TT, \Pi, \Gamma)  \models_a \psi_1) ~\wedge ((\TT, \Pi, \Gamma)  \models_a \psi_2)$} \\
$(\TT, \Pi, \Gamma) \models_a  \neg \psi$ & \text{iff} & \text{$\neg ((\TT, \Pi, \Gamma) \models_a \psi)$}  \\
$(\TT, \Pi, \Gamma) \models_a  \psi_1 \ser \psi_2$ & \text{iff} &  \begin{tabular}[c]{@{}l@{}}\text{$\exists k = \arg\min_{n \le k \le n'}, \text{for some $i \in [n, k]$}: ((\TT, \Pi, \Gamma) \models_a \psi_1)$}, for~some $j \in [k+1, n']$: \text{$((\TT, \Pi, \Gamma) \models_a \psi_2) $}\end{tabular} \\
\begin{tabular}[c]{@{}l@{}}$(\TT, \Pi, \Gamma) \models_a   [\psi]^{S,T}$\\~\end{tabular} &  \begin{tabular}[c]{@{}l@{}}\text{iff}\\~\end{tabular} & \begin{tabular}[c]{@{}l@{}} \text{$\exists k \geq n + \tau$}, s.t.~ for all $i \in [k,n+\tau']$:  $(\TT, \Pi, \Gamma) \models_a \psi$~$\wedge~ [(\Pi, \Gamma)+n' - (\Pi, \Gamma)] \in S \wedge |\Delta(\pi) - \Delta(\pi')| \in T$, \\ for some $n, n' > 0$ \end{tabular} \\ \hline
\end{tabular}
\end{table*}

\subsubsection{Asynchronous Semantics of HyperTWTL} We now introduce the notion of a trajectory, which determines when traces move and when they stutter, to model the different speeds with which traces proceed in HyperTWTL. The asynchronous semantics of HyperTWTL is the obvious choice of semantics for applications with event-driven architectures. Let us denote ${\mathcal{V}}$ as a set of trace variables. Given a HyperTWTL formula $\varphi$, we denote \textbf{Paths}$(\varphi)$ as a set of trace variables quantified in the formula $\varphi$. A given HyperTWTL formula $\varphi$ is termed $\emph{asynchronous}$ if for all propositions $a_{\pi, \rho}$ in $\varphi$, $\pi$ and $\rho$ are quantified in $\varphi$. We require that for any given formula, no trajectory or trace variable is quantified twice. A \emph{trajectory} $v:v_{0}v_{1}v_{2}\cdots$ for a given HyperTWTL formula is an infinite sequence of non-empty subsets of \textbf{Paths}($\varphi$), i.e. $v \in $ \textbf{Paths}($\varphi$). Essentially, in each step of the trajectory, one or more of the traces may progress or all may stutter. A trajectory is fair for a trace variable $\pi \in \textbf{Paths}$ if there are infinitely many positions $j$ such that $\pi \in v_{j}$. Given a trajectory $v$, by $v_{i}$, we mean the suffix $v_{i}v_{i+1}v_{i+2}\cdots$. For a set of traces variables ${\mathcal{V}}$, we denote $\mathcal{R}_{\mathcal{V}}$ as the set of all fair trajectories for indices from $\mathcal{V}$. We use trace mapping $\Pi$ as defined in the synchronous semantics of HyperTWTL. We now define the trajectory mapping $\Gamma:\textbf{Vars}(\varphi) \rightarrow \mathcal{R}_{range(\Gamma)}$, where $range(\Gamma) \subset \textbf{Vars}(\varphi)$ for which $\Gamma$ is defined. We then denote the explicit mapping of the trajectory variable $\rho$ to a trajectory $v$ as $\Gamma[\rho \rightarrow v]$. Given $(\Pi, \Gamma)$ where $\Pi$ and $\Gamma$ are the trace mapping and trajectory mapping respectively, we use $(\Pi, \Gamma) + k$ as the $k^{th}$ successor of $(\Pi, \Gamma)$. Given a trace mapping $\Pi$, a trace variable $\pi$, a trajectory variable $\rho$, a trace $\trace$, and a pointer $n$, we denote the assignment that coincides with $\Pi$ for every pair except for $(\pi, \rho)$ which is mapped to $(\trace, n)$ as $\Pi[(\pi, \rho) \rightarrow (\trace, n)]$. Given a HyperTWTL formula we denote $\Delta$ as the map from ${\mathcal{V}} \rightarrow \zplus$ that returns the time duration for each trace variable $\pi \in range(\Delta)$. For all $\pi \in range(\Delta)$, we require that the following conditions be met:
\begin{itemize}
    \item  $(\Pi, \Gamma) + k - (\Pi, \Gamma) \in S$
    \item  $|\Delta(\pi)-\Delta(\pi')| \in T$, for all distinct $\pi, \pi' \in range(\Delta)$\\
\end{itemize}   
We, therefore, present the satisfaction of asynchronous semantics of HyperTWTL formula $\varphi$ over trace mapping $\Pi$, trajectory mapping $\Gamma$, and a set of traces $\TT$ denoted as $(\TT, \Pi, \Gamma) \models_{a} \varphi$ in Table~\ref{Async}.

\begin{figure}[t!]
  \begin{center}
    \includegraphics[width=0.49\textwidth]{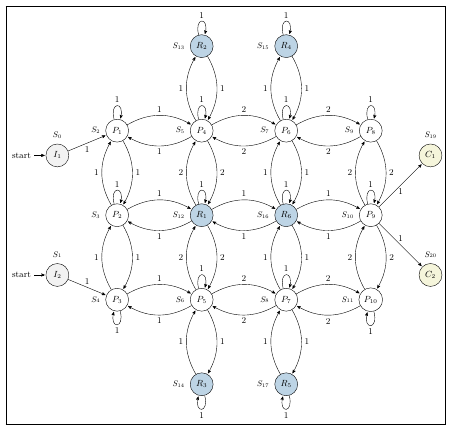}
  \end{center}
    \caption{\textcolor{black}{{TKS of the TESS case study with initial states (grey), regions of interest (blue), charging states (yellow), and allowable states (white).}}}
    \label{fig:DTS}
	\vspace{-2mm}
\end{figure}  

\section{Applications of HyperTWTL}
\label{specs}

In this section, we present a case study and illustrate some important requirements that can be expressed using HyperTWTL. This case study is inspired by the Technical Surveillance Squadron (TESS) of the United States Air Force Technical Applications Center~\cite{romano}, which provides persistent and collaborative surveillance of designated regions to detect, identify and locate potential nuclear explosions. The surveillance data is first gathered using autonomous security robots augmented with intelligent video cameras, onboard processors, communication modules, and navigation systems and then forwarded to the central control station for subsequent decision-making. 

Fig.~\ref{fig:DTS} shows the TKS of a TESS mission with different regions of interest and multiple robots. The surveillance environment is composed of $2$ initial positions $I_{1}$ and $I_{2}$ (grey), 2 charging stations $C_{1}$ and $C_{2}$ (yellow), 6 regions of interest to be surveilled $R_{1}, \dots, R_{6}$ (blue), and 10 allowable states $P_{1}, \dots, P_{10}$ (white). The weighted transitions between the states represent the possible movements of agents in the environment, with respective weights representing the unit of time required for the transitions. The autonomous surveillance can be deployed with one or multiple robots starting from any initial states $I_1$ or $I_2$ with a fully charged battery within time bound $[0, T_{1}]$. The robot(s) then proceeds along the desired routes to perform surveillance at the regions of interest. Each mission requires the surveillance of region $R_1$ within time bound $[T_{2}, T_{3}]$. Next, the robots are required to surveil either region $R_2$ followed by $R_4$ or region $R_3$ followed by $R_5$ within time bound $[T_{4}, T_{5}]$. Finally, a surveillance of region $R_6$ is performed within time bound $[T_{6}, T_{7}]$ before proceeding to a charging station $C_1$ or $C_2$ for recharging purposes within time bound $[T_{8}, T_{9}]$. Based on this case study, we consider several instances of hyperproperties that can be formalized as HyperTWTL formulae as follows. \\



\textbf{Opacity:} Information-flow security policies define what users can learn about a system while (partially) observing the system. Recent works show that robotic systems are prone to privacy/opacity attacks~\cite{chowdhury2017survey, lacava2020current}. A system is opaque if it meets two requirements: (i) there exists at least two executions of the system mapped to $\pi_1$ and $\pi_2$ with the same observations but bearing distinct secret, and (ii) the secret of each path cannot be accurately determined only by observing the system. Given a pair traces $\pi_1$  and $\pi_2$, let us assume the initial state $I$ is the only information a system user can observe, and the surveillance routes are the secret to be kept from enemy forces. Then, we observe if the assigned task is performed on both traces while having different routes but the same observations $O$. This requirement can be formalized as a HyperTWTL formula $\varphi_1$ as shown in Table~\ref{Reqs}. \\

\textbf{Non-interference:} Non-interference is a security policy that seeks to restrict the flow of information within a system. This policy requires that low-security variables be independent of high-security variables, i.e.,  one should not be able to infer information about a high-security variable by observing low-security variables. Malicious software can disrupt the communication of robotic systems and access confidential messages exchanged on the system \cite{lera2017cybersecurity}. Given any pairs of executions from the case study above, let us assume that the initial state $I$ is a high variable (high security) and paths from initial states to goal states denote a low variable (low security). The surveillance system satisfies non-interference if, there exists another execution $\pi_2$ that starts from a different high variable (i.e., the initial states are different), for all executions in $\pi_1$, and at the end of the mission, they are in the same low variable states (i.e., goal states). This requirement can be formalized as a HyperTWTL formula  $\varphi_2$ as shown in Table~\ref{Reqs}. \\

\textbf{Linearizability:} The principle underlying linearizability is that the whole system operates as if executions from all robots are from  one security robot. Thus, linearizability is a correctness condition to guarantee consistency across concurrent executions of a given system. Concurrency policies have been used to significantly enhance performance of robots applications \cite{simmons1992concurrent, rusakov2014simple}. Any pair of traces must occupy the same states within the given mission time for the surveillance mission. At the same time, it is also important to ensure that the mission's primary goal to surveil either region $R_2$ followed by $R_4$ or $R_3$ followed by $R_5$ before proceeding to the charging state $C_1$ or $C_2$ is not violated. This can be formalized as a HyperTWTL formula  $\varphi_3$ as shown in Table~\ref{Reqs}. \\

\textbf{Mutation testing}: Another interesting application of hyperproperty with quantifier alternation is the efficient generation of test cases for mutation testing. Let us assume that traces from all robots within the surveillance system are labeled as either \textit{mutated} ($\trace^{m}$) or \textit{non-mutated} ($\trace^{\neg m}$). We map $\trace^{m}$ to $\pi_{1}$  and all other non-mutated traces $\trace^{\neg m}$ to $\pi_{2}$. This requirement guarantees that even if $\pi_2$ starts from the same initial state ($I_1$ or $I_2$) as $\pi_1$, they eventually proceed to different charging states ($C_1$ or $C_2$). This can be formalized as a HyperTWTL formula  $\varphi_4$ as shown in Table~\ref{Reqs}.\\

\textbf{Side-channel timing attacks:} A side-channel timing attack is a security threat that attempts to acquire sensitive information from robotic applications by exploiting the execution time of the system. Recently, the robotic system's privacy, confidentiality, and availability have been compromised by side-channel timing attacks \cite{alahmadi2022cyber, luo2020stealthy}. To countermeasure this attack in our case study, it is required that each pair of mission execution (by a robot(s)), mapped to a pair of traces $\pi_1$ and $\pi_2$, and trajectories $\rho$ and $\rho'$, should end up in the charging state within close enough time after finishing their tasks. Let us assume that $[0, T_9]$ is the given time bound for the missions executions for $\pi_1$ and $\pi_2$, and the interval $[0,2]$ specifies how close the mission execution times should be for $\pi_1$ and $\pi_2$. This requirement can be formalized as a HyperTWTL formula  $\varphi_5$ as shown in Table~\ref{Reqs}. \\

\textbf{Observational determinism:} Observational determinism is another security policy that requires that for any given pair of traces $\pi_1$ and $\pi_2$ along the trajectory $\rho$, if the low-security inputs agree on both execution traces, then the low-security outputs must also agree in both traces. For example, given a set of executions from the case study presented above, let us assume that the initial state $I$ is a low-security input and charging state $C$ is a low-security output. The surveillance system then satisfies observation determinism if, for every pair of traces $\pi_1$ and $\pi_2$, the landing states, if the mission starts from the same initial state ($I$) for both traces, should be the same at the end of the mission within time bound $[0, T_9]$ with interval $[1,3]$ specifying how close the execution time between $\pi_1$ and $\pi_2$ should be. This requirement can be formalized as a HyperTWTL formula  $\varphi_6$ as shown in Table~\ref{Reqs}. \\

\textbf{Service level agreements:} A service level agreement (SLA) defines an acceptable performance of a given system. The expected specifications usually use statistics  such as mean response time, time service factor, percentage uptime, etc. In this requirement, for any execution with a given response time $\pi_1$, we require that there has to exist another execution $\pi_2$ along a trajectory $\rho$ with a similar timing behavior within the time bound $[T_8, T_9]$ with interval $[0,2]$ specifying how close the execution time between $\pi_1$ and $\pi_2$ should be. This requirement can be formalized as a HyperTWTL formula  $\varphi_7$ as shown in Table~\ref{Reqs}. \\

Note, all requirements in Table~\ref{Reqs} are labeled as either Synchronous (Sync) or Asynchronous (Async).
\section{Model checking of HyperTWTL}
\label{model checking}
Given a TKS $\mathcal{T}$ and a HyperTWTL formula $\varphi$, the model checking problem checks whether $\mathcal{T} \models_s \varphi$. In the later sections, we discuss the decidability of the HyperTWTL model checking problem for both the fragments, alternation-free HyperTWTL and $k$-alternation HyperTWTL, and demonstrate the method of translating an asynchronous HyperTWTL formula to a synchronous formula and translating a HyperTWTL formula into an equivalent TWTL formula. We restrict the model checking of HyperTWTL to the decidable alternation-free fragments ($\forall^{*}$- or $\exists^{*}$-) and 1-alternation fragments ($\exists^{*}\forall^{*}$). However, we do not allow fragments of HyperTWTL where a nesting structure of temporal logic formulae involves different traces. \\

\noindent \emph{Self-composition.} Let $\varphi = Q.\phi$ be a HyperTWTL formula which describes a  TKS $\mathcal{T}$, where $Q$ is a block of quantifiers and $\phi$ is the inner TWTL formula. To assert that $\mathcal{T} \models \varphi$, we generate the system $\mathcal{T'}$ which has $n$ copies of the system $\mathcal{T}$ running in parallel consisting of traces over $\mathcal{A}$. Thus, given a $2$-fold parallel self-composition of $\mathcal{T}$, we define $\mathcal{T'}$ as: $\mathcal{T'} = \mathcal{T}_{1} \times \mathcal{T}_{2} \triangleq \{(t_{0}, t'_{0}), (t_{1}, t'_{1}) \cdots \mid t \in \TT \wedge t' \in \TT \}$ \\

\noindent \textbf{Proposition 1.} Given a HyperTWTL formula $\varphi$, if there exists an equivalent TWTL formula $\varphi_{TWTL}$, then $\mathcal{T} \models_s \varphi \ratio\Leftrightarrow  \mathcal{T'} \models_s \varphi_{TWTL}$. \\


\noindent \textbf{Proof.} Let $\TT$ be a set of traces generated over the model $\mathcal{T}$, and $\TT'$ be a set of traces generated over the model $\mathcal{T'}$, so $\mathcal{T'}$ contains n copies of $\mathcal{T}$. Thus, for any set of traces $\Pi \subseteq \TT$ that satisfies the HyperTWTL formula $\varphi$, there exists a set of traces $\Pi' \subseteq \TT'$ such that $\Pi'$ satisfies the equivalent TWTL formula $\varphi_{TWTL}$, where all traces in $\Pi$ are in $\Pi'$, with unique fresh names from $\mathcal{T'}$.

\begin{algorithm}[t]
	\caption{Model checking of alternation-free HyperTWTL}
	\label{alg:dse}
		\small
        \DontPrintSemicolon

        \SetKwInOut{Input}{Inputs}\SetKwInOut{Output}{Outputs}
        \Input{HyperTWTL formula $\varphi$, TKS $\mathcal{T}$}
		\Output{Verdict = {$\bot, \top$}}

        \begin{algorithmic}[1]
        \IF{\texttt{(is\_synchronous($\varphi$))}}
            \STATE $\hat{\varphi} \leftarrow \varphi$
        \ELSE
            \STATE $\hat{\varphi} \leftarrow$ \texttt{Asynch\_to\_Synch$(\varphi)$}
        \ENDIF
        \STATE $\varphi_{TWTL} \leftarrow$ \texttt{HyperTWTL\_to\_PlainTWTL} $(\hat{\varphi})$
        \STATE $\mathcal{T}' \leftarrow$ \texttt{ModelGen}$ (\mathcal{T}, \varphi)$
        \STATE $\beta \leftarrow$ \texttt{Verify$(\mathcal{T}', \varphi_{TWTL})$}
        \RETURN $\beta$
		\end{algorithmic}
\end{algorithm}

\subsection{Model checking alteration-free HyperTWTL}
We present Algorithm 1 illustrating the overall model checking approach given a TKS $\mathcal{T}$ and an alternation-free HyperTWTL formula. The steps are described as follows.

\begin{enumerate}[label=\alph*)]
    \item First, our model checking algorithm checks if the input HyperTWTL formula $\varphi$ is synchronous or not. (Line 1)
    \item If $\varphi$ is an asynchronous formula, we translate the asynchronous HyperTWTL formula to an equivalent synchronous HyperTWTL formula using the function \texttt{Asynch\_to\_Synch()} in (Line 4), explained in detail in the next paragraph.
    \item Then, we transform the synchronous HyperTWTL formula $\hat{\varphi}$ into an equivalent TWTL formula $\varphi_{TWTL}$ using the function \texttt{HyperTWTL\_to\_PlainTWTL()}  (Line 6).
    \item Using a function \texttt{ModelGen()}, we generate a new model  that contains copies of the original model through the process of self-composition \cite{self-composition}. The number of copies equals the number of quantifiers of the formula $\varphi$ or $\psi$ (Line 7).
    \item Next, the \texttt{Verify()} function takes as inputs the new model $\mathcal{T}$' and the equivalent TWTL formula $\varphi_{TWTL}$, and then utilizes the verification approach from \cite{vasile2017time} to solve the model checking problem (Line 8).
    \item Finally, we return the verdict in case of satisfaction/violation ($\top / \bot$) (Line 9). 
\end{enumerate}

\subsection{Asynchronous HyperTWTL to Synchronous HyperTWTL} 
The process to convert a given asynchronous HyperTWTL formula to a synchronous HyperTWTL formula has two parts. First, we generate \emph{invariant} set of traces $inv(\TT)$ for the corresponding trace set $\TT$ generated over model $\mathcal{T}$. This allows for the synchronization of interleaving traces while reconciling the synchronous and asynchronous semantics of HyperTWTL. Secondly, we construct an equivalent synchronous formula $\hat{\varphi}$ from an asynchronous formula $\varphi$ such that $\TT \models_a \varphi$ if and only if $inv(\TT) \models_s \hat{\varphi}$. These steps are described as follows.

\subsubsection{Invariant Trace Generation} To construct an equivalent HyperTWTL synchronous formula $\hat{\varphi}$ from a given asynchronous HyperTWTL formula $\varphi$, we require that HyperTWTL be \emph{stutter insensitive}\cite{paviot2022structural}. To achieve this, we define the variable $\gamma_{\pi}^{\rho}$ needed for the evaluation of the atomic propositions across traces. Thus, given a pair of traces $\pi_1$ and $\pi_2$, $\gamma_{\pi}^{\rho}$ ensures that all propositions in both traces exhibit the identical sequence at all timestamps. However, since timestamps proceed at different speeds in different traces such as $\pi_1$ and $\pi_2$, a trajectory $\rho$ is used to determine which trace moves and which trace stutters at any time point. In an attempt to synchronize traces once non-aligned timestamps are identified by a trajectory, silent events ($\epsilon$) are introduced between the time stamps of the trace. For all $\trace \in \TT$, we denote \textit{inv}($\TT$) as the maximal set of traces defined over $\mathcal{A}_{\epsilon}$ where $\mathcal{A}_{\epsilon} = \mathcal{A} ~\cup~ \epsilon$. Consider a trace $\trace = (3, \{ b\})(6,\{a\})(8,\{b\})\cdots$. The trace $\trace' \in inv(\TT)$ can be generated as \textit{inv}($\trace)=\epsilon\epsilon\epsilon b \epsilon\epsilon a \epsilon b \cdots$.  We now construct the synchronous HyperTWTL formula to reason about the trace set $inv(\TT)$.

\subsubsection{Synchronous HyperTWTL Formula Construction} We now construct a synchronous formula $\hat{\varphi}$ that is equivalent to the asynchronous HyperTWTL $\varphi$. Intuitively, the asynchronous formula of HyperTWTL $\varphi$ depends on a finite interval of a timed trace. Thus, we can replace the asynchronous formula $\varphi$ with a synchronous formula $\hat{\varphi}$ that encapsulates the interval patterns in the asynchronous formula $\varphi$. Given a bounded asynchronous formula $\varphi$, we define $\beta_{\varphi}$ as the projected time period required to satisfy the asynchronous formula. Inductively, $\beta_{\varphi}$ can be defined as:
$\beta_{\HH^{d}~a}=d$ for the $\HH$ operator; $\beta_{\varphi_1 \land \varphi_2} = max(\beta_{\varphi_1} , \beta_{\varphi_2})$ for the $\land$ operator; $\beta_{\neg\varphi}=\beta_{\varphi}$ for the $\neg$ operator; $\beta_{{\varphi}_1 \ser {\varphi}_2} = \beta_{{\varphi}_1} + \beta_{{\varphi}_2}+1$ for the $\ser$ operator; $\beta_{[\varphi]^{S, T}} = up(S) + up(T)$ for the $[~]$ operator, where $up \rightarrow \zplus$ returns the upper bound of a predefined time bound. We then construct a synchronous formula $\hat{\varphi}$ from an asynchronous formula $\varphi$ by replacing the time required for the satisfaction of $\varphi$ with the appropriate $\rho_{\varphi}$. \\

\noindent \textbf{Proposition 2.} Given a set of traces $\TT$ and an alternation-free asynchronous HyperTWTL formula $\varphi$ over $\mathcal{A}$, $\TT \models \varphi$ iff \textit{inv}($\TT$) $\models_s \hat{\varphi}$.  \\ \\
\noindent Given $\TT$ is a set of traces generated over TKS, let \textit{inv}($\TT$) denote the extended set of traces for $\TT$ where $\epsilon$ has been introduced between occurrences of events to synchronize events of each $\trace' \in$ \textit{inv}($\TT$). The trace $\trace$ can be constructed from $\trace'$ by deleting the $\epsilon$-events in $\trace'$. Given the synchronous HyperTWTL formula $\hat{\varphi}$ does not violate the interval patterns of the associated asynchronous formula $\psi$, we can conclude that, if $\TT \models \varphi$, then \textit{inv}($\TT$) $\models_s \hat{\varphi}$.

\begin{table*}[!t]
\centering
\caption {\label{Reqs}Requirements expressed in HyperTWTL}
\begin{tabular}{|c|c|c|l|}
\hline
\begin{tabular}[c]{@{}c@{}}No.\\ \end{tabular} & Description & Type & \multicolumn{1}{c|}{\begin{tabular}[c]{@{}c@{}}HyperTWTL Specification\end{tabular}} \\ \hline
1 & \begin{tabular}[c]{@{}c@{}} Opacity \end{tabular}& Synch. & \begin{tabular}[c]{@{}l@{}}$\varphi_1 =\exists \pi_1 \exists \pi_2 \cdot~ [\HH^1 ~ I_{\pi_{1}} ~\wedge ~ \HH^1 ~ I_{\pi_{2}}]^{[0, T_1]} \ser~([\HH^1 ~ R_{1\pi_{1}} ~\land~ \HH^1 ~ R_{1\pi_{2}}]^{[T_2, T_3]}~\ser~([\HH^1 ~ R_{2\pi_{1}} ~\ser~ $\\$ \HH^1 ~ R_{4\pi_{1}}]^{[T_4, T_5]}~ \land~ [\HH^1 ~ R_{3\pi_{2}} ~\ser~$ $\HH^1 ~ R_{5\pi_{2}}]^{[T_4, T_5]}) \ser [\HH^1 ~ R_{6\pi_{1}} ~\land~$ $\HH^1 ~ R_{6\pi_{2}}]^{[T_6, T_7]}) $ $~\land~ $\\$[\HH^{T_{7}-T_{2}} ~ O_{\pi_{1}} =  \HH^{T_{7}-T_{2}} ~ O_{\pi_{2}}]^{[T_{2}, T_{7}]}$ \end{tabular} \\ \hline
2 & \begin{tabular}[c]{@{}c@{}}Non-\\Interference\end{tabular} & Synch. &\begin{tabular}[c]{@{}l@{}}$\varphi_2 =\exists \pi_1 \forall \pi_2 \cdot~ [\HH^1 ~ I_{\pi_{1}} ~\neq  ~ \HH^1 ~ I_{\pi_{2}}]^{[0, T_1]} ~\rightarrow~ ([\HH^1 ~ R_{1\pi_{1}} ~\land~$ $\HH^1 ~ R_{1\pi_{2}}]^{[T_2, T_3]} \ser~ ([\HH^1 ~ R_{2\pi_{1}} ~\ser~ $ \\ $ \HH^1 ~ R_{4\pi_{1}}]^{[T_4, T_5]}~\land~ [\HH^1 ~ R_{3\pi_{2}}~\ser~\HH^1 ~ R_{5\pi_{2}}]^{[T_4, T_5]}) \ser [\HH^1 ~ R_{6\pi_{1}} ~\land~$ $\HH^1 ~ R_{6\pi_{2}}]^{[T_6, T_7]}) ~\ser~ $  $[\HH^1 ~ C_{\pi_{1}} $\\$~=~ \HH^1 ~ C_{\pi_{2}}]^{[T_8, T_9]}$ \end{tabular} \\ \hline
3 & \begin{tabular}[c]{@{}c@{}}Lineariz-\\ability\end{tabular} & Synch. &\begin{tabular}[c]{@{}l@{}}$\varphi_3 =\exists \pi_1 \forall \pi_2 \cdot~ [\HH^1 ~ I_{\pi_{1}} ~= ~ \HH^1 ~ I_{\pi_{2}}]^{[0, T_1]} ~\ser~ ([\HH^1 ~ R_{1\pi_{1}} ~\land~ $ $ \HH^1 ~ R_{1\pi_{2}}]^{[T_2, T_3]} \ser~ ([\HH^1 ~ R_{2\pi_{1}} ~\ser~ $ \\ $\HH^1 ~ R_{4\pi_{1}}]^{[T_4, T_5]} \land [\HH^1 ~ R_{3\pi_{2}} ~\ser~\HH^1 ~ R_{5\pi_{2}}]^{[T_4, T_5]}) \ser [\HH^1 ~ R_{6\pi_{1}} ~\land~$ $\HH^1 ~ R_{6\pi_{2}}]^{[T_6, T_7]} \land $ \\ $[\HH^{T_{7}-T_{2}} ~ P_{\pi_{1}} ~=~ \HH^{T_{7}-T_{2}} ~ P_{\pi_{2}}]^{[T_2, T_7]})~\ser~ [\HH^1 ~ C_{\pi_{1}} ~=~ \HH^1 ~ C_{\pi_{2}}]^{[T_8, T_9]}$ \end{tabular} \\ \hline
4 & \begin{tabular}[c]{@{}c@{}}Mutation \\ Testing\end{tabular} & Synch. & \begin{tabular}[c]{@{}l@{}}$\varphi_4 =\exists \pi_1 \forall \pi_2 \cdot~ [\HH^{d} ~ t_{\pi_{1}}^{m} ~\wedge ~ \HH^d ~ t_{\pi_{2}}^{\neg m}]^{[0, T_9]} \wedge [\HH^1 ~ I_{\pi_{1}} ~= ~$ $ \HH^1 ~ I_{\pi_{2}}]^{[0, T_1]} \ser~ ([\HH^1 ~ R_{1\pi_{1}} ~\land~ $ \\ $\HH^1 ~ R_{1\pi_{2}}]^{[T_2, T_3]}~\ser~  ([\HH^1 ~ R_{2\pi_{1}} ~\ser~ \HH^1 ~ R_{4\pi_{1}}]^{[T_4, T_5]} \land $  $ [\HH^1 ~ R_{3\pi_{2}} ~\ser~$ $\HH^1 ~ R_{5\pi_{2}}]^{[T_4, T_5]}) ~\ser $ \\ $ [\HH^1 ~ R_{6\pi_{1}} ~\land~$  $\HH^1 ~ R_{6\pi_{2}}]^{[T_6, T_7]})~\ser~  [\HH^1 ~ C_{\pi_{1}} ~\neq~ \HH^1 ~ C_{\pi_{2}}]^{[T_8, T_9]}$, where $d=T_9$ \end{tabular} \\ \hline
5 & \begin{tabular}[c]{@{}c@{}}Side-Channel \\ Timing \\ Attacks\end{tabular} & Asynch. &\begin{tabular}[c]{@{}l@{}}$\varphi_5 =\forall \pi_1 \forall \pi_2 \cdot~\A \rho \E \rho' \cdot [\HH^1 ~ I_{\pi_{1}, \rho} ~\wedge ~ \HH^1 ~ I_{\pi_{2}, \rho'}]^{[0, T_1]} ~\rightarrow~ ([\HH^1 ~ R_{1\pi_{1}, \rho} ~\land~$  $ \HH^1 ~ R_{1\pi_{2}, \rho'}]^{[T_2, T_3]} \ser~ $ \\$([\HH^1 ~ R_{2\pi_{1}, \rho} ~\ser~\HH^1 ~ R_{4\pi_{1}, \rho}]^{[T_4, T_5]} \land [\HH^1 ~ R_{3\pi_{2}, \rho'} ~\ser~$  $\HH^1 ~ R_{5\pi_{2}, \rho'}]^{[T_4, T_5]}) \ser [\HH^1 ~ R_{6\pi_{1}, \rho} ~\land~  $ \\ $ \HH^1 ~ R_{6\pi_{2}, \rho'}]^{[T_6, T_7]})~\ser~[\HH^1 ~ C_{\pi_{1}, \rho} ~\land~ \HH^1 ~ C_{\pi_{2}, \rho'}]^{[T_8, T_9], [0,2]}$ \end{tabular} \\ \hline
6 & \begin{tabular}[c]{@{}c@{}}Observational\\Determinism\end{tabular} & Asynch. & \begin{tabular}[c]{@{}l@{}}$\varphi_6 = \exists \pi_2 \forall \pi_1 \cdot~ \A \rho\cdot[\HH^1 ~ I_{\pi_{1}, \rho} ~= ~ \HH^1 ~ I_{\pi_{2}, \rho}]^{[0, T_1]} ~\rightarrow~ ([\HH^1 ~ R_{1\pi_{1},\rho} ~\land~$  $\HH^1 ~ R_{1\pi_{2},\rho}]^{[T_2, T_3]} \ser~ $\\ $([\HH^1 ~ R_{2\pi_{1},\rho} ~\ser~\HH^1 ~ R_{4\pi_{1},\rho}]^{[T_4, T_5]} \land [\HH^1 ~ R_{3\pi_{2}, \rho}~\ser~ $  $\HH^1 ~ R_{5\pi_{2}, \rho}]^{[T_4, T_5]}) \ser [\HH^1 ~ R_{6\pi_{1}, \rho} ~\land~$  \\ $\HH^1 ~ R_{6\pi_{2}, \rho}]^{[T_6, T_7]}) ~\ser~ [\HH^1 ~ C_{\pi_{1}, \rho} ~=~ \HH^1 ~ C_{\pi_{2}, \rho}]^{[T_8, T_9][1,3]}$ \end{tabular} \\ \hline
7 & \begin{tabular}[c]{@{}c@{}}Service Level \\Agreement \end{tabular} & Asynch. & \begin{tabular}[c]{@{}l@{}} {~}\\ $\varphi_7 = \exists \pi_2 \forall \pi_1 \cdot~ \E \rho\cdot[\HH^1 ~ I_{1\pi_{1}, \rho} ~\land~ \HH^1 ~ I_{1\pi_{2}, \rho}]^{[0, T_1]} ~\rightarrow~ [\HH^1 ~ C_{\pi_{1},\rho} ~\land~ $ $ \HH^1 ~ C_{\pi_{2},\rho}]^{[T_8, T_9], [0,2]}$\\{~} \end{tabular} \\  \hline

\end{tabular}
\end{table*}

\begin{table*}[!t]
\centering
\caption{\label{equi-TWTL} Equivalent TWTL formulae of HyperTWTL in Table~\ref{Reqs}} 
\begin{tabular}{|c|l|}
\hline
No. & \multicolumn{1}{c|}{TWTL Specifications} \\ \hline
1 & \begin{tabular}[c]{@{}l@{}}$\theta_{1} =  ([\HH^1 M_{\phi_1}^{1}]^{[0, T_1]} ~\ser~ [\HH^{1} M_{\phi_2}^{1}]^{[T_{2},T_{3}]} ~\ser~ ([\HH^{1} R_{2}^{1} \ser \HH^{1} R_{4}^{1}]^{[T_4, T_5]} \lor [\HH^{1} R_{3}^{1} ~\ser $ $\HH^{1} R_{5}^{1}]^{[T_4, T_5]}) \ser [\HH^{1} M_{\phi_3}^{1}]^{[T_6, T_7]}  $\\$\land~ [\HH^{T_{2}-T_{7}} M_{\phi_4}^{1}]^{[T_7, T_2]}) \land ([\HH^1 M_{\phi_1}^{2}]^{[0, T_1]} \ser [\HH^{1} M_{\phi_2}^{2}]^{[T_{2},T_{3}]} \ser ([\HH^{1} R_{2}^{2} ~\ser \HH^{1} R_{4}^{2}]^{[T_4, T_5]} \lor [\HH^{1} R_{3}^{2}$  $\ser ~\HH^{1} R_{5}^{2}]^{[T_4, T_5]})~ \ser $\\$ [\HH^{1} M_{\phi_3}^{2}]^{[T_6, T_7]} \land [\HH^{T_{7}-T_{2}} M_{\phi_4}^{2}]^{[T_2, T_7]})$\end{tabular} \\ \hline
2 & \begin{tabular}[c]{@{}l@{}}$\theta_{2} = ([\HH^1 I_{\psi_1}^{1}]^{[0, T_1]} \rightarrow [\HH^1 B_{\psi_2}^{1}]^{[T_2, T_3]} \ser ([\HH^1 R_{2}^{1} \ser \HH^1 R_{4}^{1}]^{[T_4, T_5]} \lor [\HH^1 R_{3}^{1} ~\ser$  $\HH^1 R_{5}^{1}]^{[T_4, T_5]}) \ser [\HH^1 B_{\psi_3}^{1}]^{[T_{6}, T_{7}]} \ser$ $[\HH^1 B_{\psi_4}^{1}]^{[T_8, T_9]}) $\\$\land~ ([\HH^1 B_{\psi_1}^{2}]^{[0, T_1]} \rightarrow$ $ [\HH^1 B_{\psi_2}^{2}]^{[T_2, T_3]} \ser ([\HH^1 R_{2}^{2} \ser \HH^1 R_{4}^{2}]^{[T_4, T_5]} \lor [\HH^1 R_{3}^{2} \ser \HH^1 R_{5}^{2}]^{[T_4, T_5]}) ~\ser [\HH^1 B_{\psi_3}^{2}]^{[T_6, T_7]} $  $ \ser [\HH^1 B_{\psi_4}^{2}]^{[T_8, T_9]})$\end{tabular} \\ \hline
3 & \begin{tabular}[c]{@{}l@{}}$\theta_{3} = ([\HH^1 B_{\psi_1}^{1} ]^{[0, T_1]} \ser [\HH^1 B_{\psi_2}^{1}]^{[T_2, T_3]} \ser ([\HH^1 R_{2}^{1} \ser \HH^1 R_{4}^{1}]^{[T_4, T_5]} \lor [\HH^1 R_{3}^{1} ~\ser$ $\HH^1 R_{5}^{1}]^{[T_4, T_5]}) \ser [\HH^1 B_{\psi_3}^{1}]^{[T_6, T_7]} \ser~ [\HH^1 B_{\psi_4}^{1}]^{[T_8, T_9]}$ \\ $ \land [\HH^{T_{7}-T_{2}} B_{\psi_5}^{1}]^{[T_2, T_7]}) \lor  ([\HH^1 B_{\psi_1}^{2}]^{[0, T_1]} \ser [\HH^1 B_{\psi_2}^{2}]^{[T_2, T_3]} \ser ([\HH^1 R_{3}^{2} ~\ser$ $\HH^1 R_{5}^{2}]^{[T_4, T_5]} \lor [\HH^1 R_{3}^{2} ~\ser~  \HH^1 R_{5}^{2}]^{[T_4, T_5]})~ \ser~$\\$  [\HH^1 B_{\psi_3}^{2}]^{[T_6, T_7]}~ \ser [\HH^1 B_{\psi_4}^{2}]^{[T_8, T_9]} \land [\HH^{T_{7}-T_{2}} B_{\psi_5}^{2}]^{[T_2, T_7]})$\end{tabular} \\ \hline
4 &  \begin{tabular}[c]{@{}l@{}}$\theta_{3} = ([\HH^1 B_{\psi_1}^{1} ]^{[0, T_1]} \wedge [\HH^1 B_{\psi_2}^{1}]^{[T_2, T_3]} \ser ([\HH^1 R_{2}^{1} \ser \HH^1 R_{4}^{1}]^{[T_4, T_5]} \lor [\HH^1 R_{3}^{1} ~\ser$ $\HH^1 R_{5}^{1}]^{[T_4, T_5]}) \ser [\HH^1 B_{\psi_3}^{1}]^{[T_6, T_7]} \ser$  $[\HH^1 B_{\psi_4}^{1}]^{[T_8, T_9]}) \lor $\\$ ([\HH^1 B_{\psi_1}^{2}]^{[0, T_1]} \ser [\HH^1 B_{\psi_2}^{2}]^{[T_2, T_3]} \ser ([\HH^1 R_{2}^{2} ~\ser$ $\HH^1 R_{4}^{2}]^{[T_4, T_5]} \lor [\HH^1 R_{3}^{2} ~\ser  $ $  \HH^1 R_{5}^{2}]^{[T_4, T_5]})~ \ser~[\HH^1 B_{\psi_3}^{2}]^{[T_6, T_7]} \ser [\HH^1 B_{\psi_4}^{2}]^{[T_8, T_9]} )$\end{tabular} \\ \hline
5 & \begin{tabular}[c]{@{}l@{}}$\theta_{5} = ([\HH^1 M_{\phi_1}^{1} ]^{[0, T_1]} \rightarrow [\HH^1 M_{\phi_2}^{1}]^{[T_2, T_3]} \ser ([\HH^1 R_{2}^{1} \ser \HH^1 R_{4}^{1}]^{[T_4, T_5]} \lor [\HH^1 R_{3}^{1} ~\ser$ $\HH^1 R_{5}^{1}]^{[T_4, T_5]}) \ser [\HH^{1} M_{\phi_3}^{1}]^{[T_{6}, T_{7}]} ~\ser~ [\HH^{1} M_{\phi_4}^{1}]^{[T_{8}, T_{9}]}) $\\$ \land ([\HH^1 M_{\phi_1}^{2} ]^{[0, T_1]} \rightarrow $ $ [\HH^1 M_{\phi_2}^{2}]^{[T_2, T_3]} \ser ~([\HH^1 R_{2}^{2} \ser \HH^1 R_{4}^{2}]^{[T_4, T_5]} \lor [\HH^1 R_{3}^{2} ~\ser$ $~\HH^1 R_{5}^{2}]^{[T_4, T_5]}) ~\ser $ $ [\HH^{1} M_{\phi_3}^{2}]^{[T_{6}, T_{7}]} \ser [\HH^{1} M_{\phi_4}^{2}]^{[T_{8}, T_{9}]})$\end{tabular} \\ \hline
6 & \begin{tabular}[c]{@{}l@{}}$\theta_{6} = ([\HH^1 B_{\psi_1}^{1} ]^{[0, T_1]} \rightarrow [\HH^1 B_{\psi_2}^{1}]^{[T_2, T_3]} \ser ([\HH^1 R_{2}^{1} \ser \HH^1 R_{4}^{1}]^{[T_4, T_5]} \lor [\HH^1 R_{3}^{1} ~\ser$  $\HH^1 R_{5}^{1}]^{[T_4, T_5]}) \ser [\HH^1 B_{\psi_3}^{1}]^{[T_{6}, T_{7}]} \ser$ $[\HH^1 B_{\psi_4}^{1}]^{[T_8, T_9]}) $\\$\land~([\HH^1 B_{\psi_1}^{2}]^{[0, T_1]} \rightarrow$ $ [\HH^1 B_{\psi_2}^{2}]^{[T_2, T_3]} \ser ([\HH^1 R_{2}^{2} \ser \HH^1 R_{4}^{2}]^{[T_4, T_5]} \lor [\HH^1 R_{3}^{2} \ser \HH^1 R_{5}^{2}]^{[T_4, T_5]}) ~\ser $  $[\HH^1 B_{\psi_3}^{2}]^{[T_6, T_7]}  \ser [\HH^1 B_{\psi_4}^{2}]^{[T_8, T_9]})$\end{tabular} \\ \hline
7 & \begin{tabular}[c]{@{}l@{}}{~}\\ $\theta_{7} = ([\HH^1 M_{\phi_1}^{1} ]^{[0, T_1]} \rightarrow [\HH^1 M_{\phi_2}^{1}]^{[T_8, T_9]}) \land ([\HH^1 M_{\phi_1}^{2} ]^{[0, T_1]} \rightarrow $ $ [\HH^1 M_{\phi_2}^{2}]^{[T_8, T_9]})$ \\ {~}\end{tabular} \\\hline

\end{tabular}
\end{table*}

\subsection{Converting HyperTWTL to TWTL formula} 
We verify the HyperTWTL specifications by creating a new model that contains copies of the original system, where the number of copies is equal to the number of quantifiers in the HyperTWTL formula. The new model is then verified against a given HyperTWTL specification. 
Given a HyperTWTL formula $\varphi$, let $\phi_{1} \dots \phi_{N \in \zplus}$ be the sub-formulae of $\varphi$, where the same proposition is observed on any pair of traces. We denote $M_{\phi_{i}}^{j}$ as an instance of $\phi_{i}$ where $i \leq N$ and $j \in \zplus$ is the index of the copy of the original model. For example, consider the HyperTWTL formula $\varphi_1$ in Table~\ref{Reqs}. The following sub-formulae of $\varphi_1$ analyze the same proposition on the given pair of traces $\pi_1$ and $\pi_2$: $\phi_1 = [\HH^1 ~ I_{\pi_{1}} ~\wedge ~ \HH^1 ~ I_{\pi_{2}}]^{[0, T_1]}$, $\phi_2 = [\HH^1 ~ R_{1\pi_{1}} ~\land~ \HH^1 ~ R_{1\pi_{2}}]^{[T_2, T_3]}$, $\phi_3 = [\HH^1 ~ R_{6\pi_{1}} ~\land~\HH^1 ~ R_{6\pi_{2}}]^{[T_6, T_7]}$, and $\phi_4 = [\HH^{T_{7}-T_{2}} ~ O_{\pi_{1}} =  \HH^{T_{7}-T_{2}} ~ O_{\pi_{2}}]^{[T_{2}, T_{7}]}$
We obtain an equivalent TWTL formula by removing the quantifier prefix and introducing fresh atomic propositions that capture the notion of occurrences of the observation of the same proposition on any pair of traces. In the case where different propositions are to be observed on a given pair of traces, we maintain the proposition while introducing a superscript $j \in \zplus$ for the same purpose as described above. Given $\varphi_1$ has two quantifiers, two copies of the original model will be needed to verify the equivalent TWTL formula. Thus, we denote $M_{\phi_{1}}^{1}$ and $M_{\phi_{1}}^{2}$ as instances of $\phi_1$ for the first and second copies of the model. Similarly, $M_{\phi_{3}}^{1}$ and $M_{\phi_{3}}^{2}$ are the first and second copies of the model for the instance of $\phi_3$. The HyperTWTL formula $\varphi_1$ can then be translated as an equivalent TWTL as $\theta_{1} =  ([\HH^1 M_{\phi_1}^{1}]^{[0, T_1]} ~\ser~ [\HH^{1} M_{\phi_2}^{1}]^{[T_{2},T_{3}]} ~\ser~ ([\HH^{1} R_{2}^{1} \ser$ $\HH^{1} R_{4}^{1}]^{[T_4, T_5]} \lor [\HH^{1} R_{3}^{1} ~\ser \HH^{1} R_{5}^{1}]^{[T_4, T_5]}) \ser [\HH^{1} M_{\phi_3}^{1}]^{[T_6, T_7]}$ $\land [\HH^{T_{7}-T_{2}} M_{\phi_4}^{1}]^{[T_7, T_2]}) \land ([\HH^1 M_{\phi_1}^{2}]^{[0, T_1]} ~\ser [\HH^{1} M_{\phi_2}^{2}]^{[T_{2},T_{3}]}$  $\ser ([\HH^{1} R_{2}^{2} ~\ser \HH^{1} R_{4}^{2}]^{[T_4, T_5]} \lor [\HH^{1} R_{3}^{2} \ser$ \\$ ~\HH^{1} R_{5}^{2}]^{[T_4, T_5]}) ~\ser$ $[\HH^{1} M_{\phi_3}^{2}]^{[T_6, T_7]} \land [\HH^{T_{7}-T_{2}} M_{\phi_4}^{2}]^{[T_7, T_2]}).$ \\

In the case of $\varphi_5$ (since it is an asynchronous formula), we first translate the formula into an equivalent synchronous HyperTWTL formula using the method we described in Section \ref{model checking}(B). The resulting synchronous formula is then translated into an equivalent TWTL formula using the same method described in the previous example of $\varphi_1$ and is shown in Table~\ref{equi-TWTL}. Compared to $\varphi_1$ and $\varphi_5$, the translation of $\varphi_2$, $\varphi_3$, $\varphi_4$, $\varphi_6$, and $\varphi_7$  to equivalent TWTL formulae is not straightforward. This is because the model checking problem of HyperTWTL approaches undecidability when only a single quantifier alternation is allowed \cite{ho2018verifying}. However, model checking of HyperTWTL formulae of the form  $\exists^{*}\forall^{*}$ is decidable when the universal quantifier is flattened. For instance, consider the HyperTWTL formula $\varphi_3$ in Table \ref{Reqs}. To solve this formula, we flatten the literals associated with $\pi_{2}$ by enumerating all the possible interactions between $\pi_{1}$ and $\pi_2$, thus reducing the problem to a $\exists^{*}\exists^{*}$ problem. Let us denote $\psi_{1}\dots\psi_{N\in \zplus}$ as sub-formulae of $\varphi_3$ which encapsulates all the possible interactions between $\pi_1$ and $\pi_2$. Thus, the first sub-formula of $\varphi_3$, $[H^{1} I_{\pi_{1}} = H^{1} I_{\pi_{2}}]^{[0, T_{1}]}$ yields $\psi_{1} = [H^{1} I_{1\pi_{1}} \land H^{1} I_{2\pi_{2}}]^{[0, T_{1}]} \lor [H^{1} I_{2\pi_{1}} \land H^{1} I_{1\pi_{2}}]^{[0, T_{1}]}$. We then denote $B_{\psi_{i}}^{j}$ as an instance of $\psi_{i}$ where $i \leq N$ and $j \in \zplus$ is the index of the copy of the original model. Once the alternation is eliminated, $\varphi_3$ and $\varphi_4$ can be converted to equivalent TWTL formulae using the same approach described for $\varphi_1$ and $\varphi_2$ as shown in Table~\ref{equi-TWTL}.

\subsection{Model checking $k$-alternation HyperTWTL}

Model checking $k$-alternations formulae are generally complex as all given executions have to be examined. For instance, consider the HyperTWTL formula $\varphi = \exists \pi_1.\forall \pi_2 \cdot \phi$. To verify this formula, it requires that for all traces $\trace \in \TT$, there exists a trace $\trace$ that the formula $\phi$ is violated. The situation is dire in specifications with more than one alternation of quantifiers. Model checking such specifications may lead to potential state explosion, even with a finite set of traces. Consistent with the general notion of undecidability of a model checking problem, $\exists^* \forall^*$ and $\forall^* \exists^*$ fragments of HyperTWTL are undecidable in both the synchronous and asynchronous semantics. Despite the difficulty and complexity of model checking of HyperTWTL beyond the alternation-free fragments, $k$-alternation fragment of HyperTWTL can be decided within a  bounded time domain as follows. 

The model checking for HyperTWTL becomes decidable when there is an \textit{a priori} bound $k_{lim}$ on the variability of the traces. The variability of a timed trace is the maximum possible number of events in any open unit interval. Hence, given the bound $k_{lim} \in \zplus$, the number of events in a timed trace is less than or equal to $k_{lim}$. Therefore, the verification of synchronous HyperTWTL (at least for the $\exists^*\forall^*$-fragment) can be decided with any tool that works with TWTL. With asynchronous HyperTWTL $\psi$, given a set of traces $\TT$ and a variability bound $k_{lim}$, we only evaluate traces whose timestamps are less than $k_{lim}$. Let $\TT_{[0, k_{lim})}$ denote the set of all such traces in $\TT$, i.e. $\TT_{[0, k_{lim})} \subseteq \TT$. We can now reduce the model checking problem to $\TT_{[0, k_{lim})} \models_s \psi$.  \\

\noindent \textbf{Proposition 3.} Model checking HyperTWTL can be decided when all traces have constrained variability $k_{lim}$ where $k_{lim} \in \zplus$.

\subsection{Complexity}
In this section, we review the complexity of Algorithm 1 for the model checking of alternation-free HyperTWTL formulae. The time complexity of translating an asynchronous HyperTWTL to synchronous HyperTWTL formula is based on the structure of the formula. Translating the asynchronous formula to a synchronous formula in Algorithm 1 takes $O(|\varphi|)$ at most. The time complexity of translating HyperTWTL to TWTL is upper-bounded to $O(|\varphi|\cdot 2^{|AP|})$. The model generation function in Algorithm 1 depends on the structure of the formula and the number of quantifiers in the given formula. The time complexity of the model generation is $O(|Q|^n)$, where $n$ is the number of copies of the original model. The satisfiability problem of HyperTWTL can be solved similarly as for HyperLTL, i.e., satisfiability is decidable for HyperTWTL fragments not containing $\forall\exists$ irrespective of the semantics used. Hence, the complexity results for the various fragments of HyperTWTL will then be similar to that of HyperLTL, i.e., the satisfiability problem for the alternation-free fragment and bounded $\exists^{*}\forall^{*}$ fragments of HyperTWTL is thus PSPACE-complete.

\section{Implementation and Results}
To illustrate the effectiveness of our proposed methods, in this section, we present the evaluation of the TESS case study described in Section~\ref{specs} using Algorithm 1. The HyperTWTL specifications for this case study are presented in Table~\ref{Reqs} where $\varphi_{1} - \varphi_{4}$ are synchronous formulae and  $\varphi_5 - \varphi_{7}$ are asynchronous formulae. All formulae were converted into equivalent TWTL formulae (see Section \ref{model checking}) as shown in Table \ref{equi-TWTL}. 
For verifying the formulae in Table \ref{equi-TWTL}, we use the PyTWTL tool~\cite{twtltool}, a Python 2.7 implementation of the TWTL verification algorithms proposed in \cite{vasile2017time}. All the experiments are performed on a Windows 10 system with 64 GB RAM and Intel Core(TM) i9-10900 CPU (3.70 GHz). The following time bounds are considered for the verification of all the equivalent flat TWTL properties in Table~\ref{equi-TWTL}:  $T_1 = 2, ~T_2 = 5,~ T_3 = 6,~ T_4 = 7,~ T_5 = 12,~ T_6 = 13,~ T_7 = 19,~ T_8 = 20$ and $T_9 = 30$. The obtained verification results are shown in Table~\ref{results-TWTL}. We observe that the case study satisfies the specification TWTL specification $\theta_1$, $\theta_5$, and $\theta_7$. 
However, the rest of the formulae $\theta_2$, $\theta_3$, $\theta_4$, and $\theta_6$ were unsatisfied. This means that not all the runs over $\mathcal{T'}$ satisfy those formulae. For instance, let us consider the case of $\theta_2$. In $\mathcal{T'}$, there exists atleast at least one path, such as $(I_2 \rightarrow P_3 \rightarrow P_5 \rightarrow R_1 \rightarrow P_4 \rightarrow R_2 \rightarrow P_4 \rightarrow P_6 \rightarrow R_4 \rightarrow P_6 \rightarrow R_6 \rightarrow P_9 \rightarrow C_1)  \wedge (I_1 \rightarrow P_1 \rightarrow P_4 \rightarrow R_1 \rightarrow P_5 \rightarrow R_3 \rightarrow P_5 \rightarrow P_7 \rightarrow R_5 \rightarrow P_7 \rightarrow R_6 \rightarrow P_9 \rightarrow C_2)$ over $\mathcal{T'}$ that violates $\theta_2$. We observe that if the aforementioned path is mapped to the trace variables $\pi_1$ and $\pi_2$ respectively, the pair of traces also violate the HyperTWTL specification $\varphi_2$. This shows that a trace over $\mathcal{T'}$ that violates TWTL formula is equivalent to a pair of traces over $\mathcal{T}$ that violates the corresponding HyperTWTL formula.
We also observe that the execution time for verifying the TWTL specifications $\theta_1, \theta_2, \theta_3, \theta_4, \theta_5, \theta_6$, and $\theta_7$ are $16.30$, $19.05$, $22.24$, $24.52$, $25.19$, $23.83$, and $22.24$ seconds, respectively. Similarly, the memory consumed for verifying these specifications are $15.21$, $15.33$, $17.25$, $16.27$, $16.04$, $16.39$, and $17.48$ MB, respectively. 


\begin{table}[!t]
\centering
\caption{\label{results-TWTL} Experimental results for model checking TWTL formulae in Table \ref{equi-TWTL} } 
\begin{tabular}{|c|c|c|c|}
\hline
\begin{tabular}[c]{@{}c@{}}TWTL\\ Specification\end{tabular} & Verdict & \begin{tabular}[c]{@{}c@{}}Time\\ (Seconds)\end{tabular} & \begin{tabular}[c]{@{}c@{}}Memory\\ (MB)\end{tabular} \\ \hline
$\theta_{1} $ & SAT & 16.30 & 15.21 \\ \hline
$\theta_{2} $ & UNSAT & 19.05 & 15.33 \\ \hline
$\theta_{3} $ & UNSAT & 22.24 & 17.25 \\ \hline
$\theta_{4} $ & UNSAT & 24.52 & 16.27 \\ \hline
$\theta_{5} $ & SAT & 25.19 & 16.04 \\ \hline
$\theta_{6} $ & UNSAT & 23.83 & 16.39 \\ \hline
$\theta_{7} $ & SAT & 22.24 & 17.48 \\ \hline
\end{tabular}
\end{table}

In addition to model checking of HyperTWTL, our proposed approach can be applied to other applications. One such application is motion planning in robotics.  
Let us assume a given 2-D grid representation where the green, red, grey, blue, and white colors represent the initial, goal, region of interest, obstacles, and the allowable states respectively. We formalize two planning objectives using HyperTWTL on the 2-D grid as follows. \\

\textbf{Requirement 8 (Shortest path):} The shortest path exists if there exists a trace $\pi_2$ that starts from the initial state through the region of interest and reaches the goal state before any other trace $\pi_1$. We consider the following time bounds for the synthesis of this requirement: $T_1 = 2, T_2 = 3, T_3 = 8, T_4 =9, T_5 = 13$. This requirement can be formalized as a HyperTWTL formula as: \\

$\varphi_8 = \exists \pi_2 \forall \pi_1 \cdot~ [\HH^1 ~ s_{{0}_{{\pi}_1}}$
~=~ $\HH^1 ~ s_{{0}_{{\pi}_2}}]^{[0, T_1]} ~\ser~ [\HH^1 ~ r_{{\pi}_1} ~\land~$ $ \HH^1 ~ r_{{\pi}_2}]^{[T_2, T_3]} ~\ser~([\HH^1 ~ g_{{\pi}_2}]^{[T_{4}, T_5]} ~\land [\HH^1  g_{{\pi}_2}] \rightarrow [\HH^1  g_{{\pi}_1}]^{[T_4, T_5]})$ \\

\textbf{Requirement 9 (Initial-state opacity):} The opacity requirement is satisfied when at least two traces $\pi_1$ and $\pi_2$ meet the following conditions : (i) both traces of the system mapped to $\pi_1$ and $\pi_2$ have the same observations but bear different secrets; and (ii) the secret of each path cannot be accurately determined by observing(partially) the system alone. For initial-state opacity requirement, let the initial state of the paths ($s_{0} \in S_{init}$) be secret and observe whether both paths reach the goal states with the same set of observations. We consider the following time bounds for the synthesis of this requirement: $T_1 = 2s, T_2 = 3s, T_3 = 8s, T_4 =9s, T_5 = 13s$. This objective can be expressed using HyperTWTL as: \\

 $\varphi_9 =\exists \pi_1 \exists \pi_2 \cdot ~[\HH^1 ~ s_{{0}_{{\pi}_1}} ~\neq~ \HH^1 ~ s_{{0}_{{\pi}_2}}]^{[0, T_1]} ~\ser~ [\HH^1 ~ r_{{\pi}_1} ~\land~ $ ~$ \HH^1 ~ r_{{\pi}_2}]^{[T_2, T_3]} ~\ser~ [\HH^1 ~ g_{{\pi}_1} ~\land~ \HH^1 ~ g_{{\pi}_2}]^{[T_4, T_5]} $ \\



\begin{figure}[!t]
    \centering
    {\includegraphics[width=0.48\linewidth]{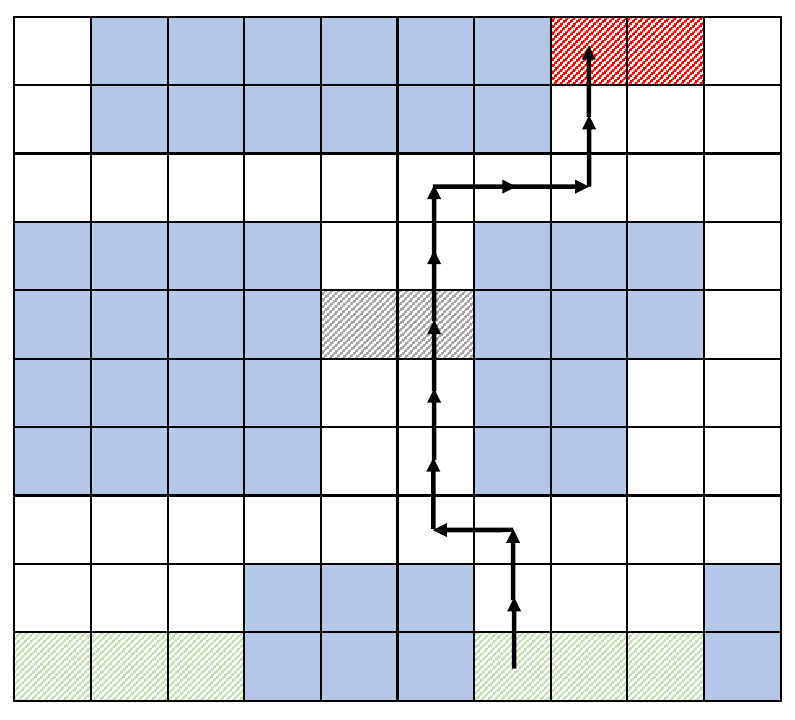}}
           \caption{Strategy for shortest path $\varphi_5$}
        \label{fig:graphs1}
\end{figure}

\begin{figure}[!t]
    \centering
    {\includegraphics[width=0.48\linewidth]{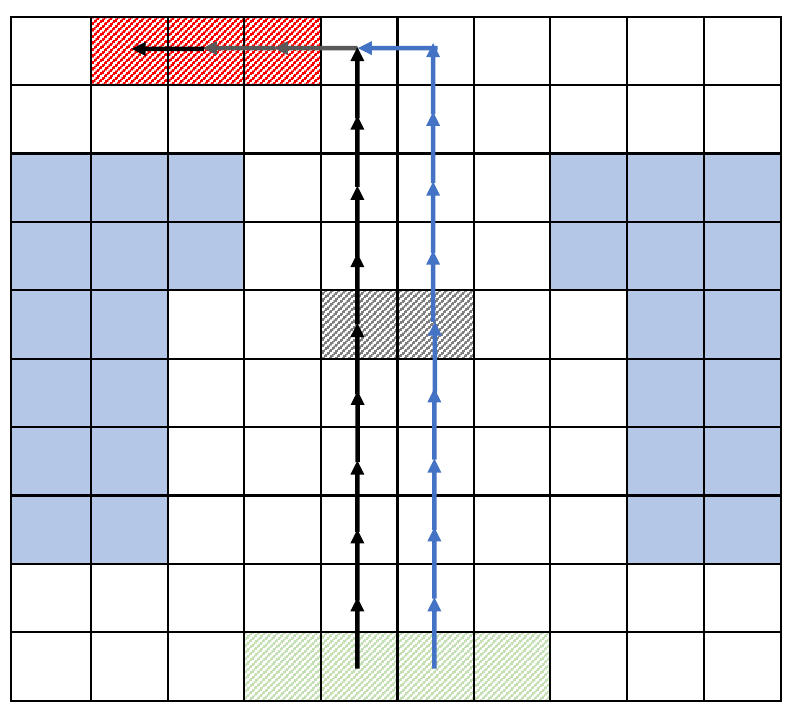}}
           \caption{Strategy for initial-state opacity $\varphi_6$}
        \label{fig:graphs2}
\end{figure}

The 2-D grid environment is first converted into an equivalent TKS, and then fed to Algorithm 1 with each planning objective formalized as $\varphi_8$ and $\varphi_9$. For this, we make use of the \texttt{Synthesis()} function from \cite{vasile2017time} instead of the \texttt{Verify()} function in Algorithm 1 (Line 8) to solve the planning problem using the PyTWTL tool.
For the shortest path objective, $\varphi_8$, the synthesized black path in Figure \ref{fig:graphs1} is the shortest path from the initial state through to the region of interest to the goal state. In Figure \ref{fig:graphs2}, the synthesized path in black is the feasible strategy for the initial-state opacity objective that has the same observation as the path in blue. Both paths reach the goal state despite starting from a different initial state. \\

Next, we evaluate the scalability and performance of the proposed approach for path synthesis. We used varying grid sizes ranging from $10 \times 10$ to $50 \times 50$ to synthesize paths for the formalized HyperTWTL specifications $\varphi_8$ and $\varphi_9$.  We then analyze the impact of the increasing sizes on the performance of the proposed tool. The respective synthesis time and memory consumed are shown in Table~\ref{Syn_time}. We observe from Table~\ref{Syn_time} that the time taken for our algorithm to synthesize a path increases with an increase in the size of the grid. For instance, the algorithm takes $23.08$ seconds to synthesize a path for $\varphi_8$ on a 10$\times$10 grid. However, while synthesizing a feasible path for the same planning objective on a grid size of 20$\times$20, the synthesis time increases to $47.94$ seconds. Similarly, the synthesis time increases to $72.40$ seconds,  $103.05$ seconds, and $161.17$ seconds while synthesizing $\varphi_8$ on 30$\times$30, 40$\times$40, and 50$\times$50 grid sizes respectively. Again, while synthesizing a feasible path for $\varphi_9$ on a 10$\times$10 grid size, our algorithm takes $17.43$ seconds. The synthesis time increases to $21.63$ seconds while synthesizing a feasible path for the same objective on a grid size 20$\times$20. Once again, the synthesis time increases to $28.61$ seconds,  $36.72$ seconds, and $56.49$ seconds while synthesizing $\varphi_8$ on 30$\times$30, 40$\times$40, and 50$\times$50 grid sizes respectively. Consequently, as shown in Table~\ref{Syn_time}, we observe that the memory consumed by our algorithm increases with an increase in the grid size. For instance, the algorithm consumes $15.53$ MB to synthesize a path for $\varphi_8$ on a 10$\times$10 grid size. Again, while synthesizing a feasible path for the same specification on a grid size of 20$\times$20, the memory consumed increases to $19.11$ MB. Similarly, the memory consumed increases to $28.84$ MB,  $39.50$ MB, and $52.11$ seconds while synthesizing $\varphi_8$ on 30$\times$30, 40$\times$40 and 50$\times$50 grid sizes respectively. Again, our algorithm consumes $15.21$ MB while synthesizing a feasible path for $\varphi_9$ on a 10$\times$10 grid size.  While synthesizing a feasible path for the same objective on a grid size 20$\times$20, the memory consumed increases to $18.96$ MB. Once again, the memory consumed increases to $23.73$ MB,  $27.35$ MB, and $31.96$ MB while synthesizing $\varphi_8$ on 30$\times$30, 40$\times$40, and 50$\times$50 grid sizes, respectively.

\begin{table}[t]
\centering
\caption {Comparison of synthesis times and memory consumed for HyperTWTL planning objectives $\varphi_8$ and $\varphi_9$}\label{Syn_time}
\begin{tabular}{|c|l|c|c|}
\hline
\begin{tabular}[c]{@{}c@{}}HyperTWTL\\ Specification\end{tabular} & \multicolumn{1}{c|}{\begin{tabular}[c]{@{}c@{}}Grid\\ size\end{tabular}} & \begin{tabular}[c]{@{}c@{}}Time\\ (Seconds)\end{tabular} & \begin{tabular}[c]{@{}c@{}}Memory\\ (MB)\end{tabular} \\ \hline
$\varphi_{8} $ & \multirow{2}{*}{10$\times$10} & 23.08 & 15.53 \\ \cline{1-1} \cline{3-4} 
$\varphi_{9} $ &  & 17.43 & 15.21 \\ \hline
$\varphi_{8} $ & \multicolumn{1}{c|}{\multirow{2}{*}{20$\times$20}} & 47.94 & 19.11 \\ \cline{1-1} \cline{3-4} 
$\varphi_{9} $ & \multicolumn{1}{c|}{} & 21.63 & 18.96 \\ \hline
$\varphi_{8} $ & \multirow{2}{*}{30$\times$30} & 72.40 & 28.84 \\ \cline{1-1} \cline{3-4} 
$\varphi_{9} $ &  & 28.61 & 23.73 \\ \hline
$\varphi_{8} $ & \multirow{2}{*}{40$\times$40} & 103.05 & 39.50 \\ \cline{1-1} \cline{3-4} 
$\varphi_{9} $ &  & 36.72 & 25.35 \\ \hline
$\varphi_{8} $ & \multirow{2}{*}{50$\times$50} & 161.17 & 52.11 \\ \cline{1-1} \cline{3-4} 
$\varphi_{9} $ &  & 56.49 & 31.96 \\ \hline

\end{tabular}
\end{table}

\section{Related Works}
\label{related-works}
HyperLTL and HyperCTL$^{*}$ which were first introduced in \cite{clarkson2014temporal} extend the temporal logics LTL, CTL, and CTL$^{*}$ with explicit and concurrent quantifications over trace executions of a system. In recent times, multiple techniques have been proposed to monitor \cite{brett2017rewriting, agrawal2016runtime, bonakdarpour2018monitoring, stucki2019gray} and verify \cite{coenen2019verifying, finkbeiner2020synthesis, finkbeiner2017verifying} hyperproperties expressed as HyperLTL and HyperCTL$^{*}$ specifications. Similarly, other techniques have been proposed to monitor other hyper-temporal logics. HyperSTL is a bounded hyper-temporal logic for specifying hyperproperties over real-valued signals. A testing technique for verifying HyperSTL properties in cyber-physical systems is proposed in \cite{nguyen2017hyperproperties}. This testing technique allows for the falsification or checking of bounded hyperproperties in CPS models. HyperMTL, a hyper-temporal logic that addresses some limitations of HyperLTL in formalizing bounded hyperproperties, is proposed in \cite{bonakdarpour2018opportunities}. 
In \cite{ho2018verifying}, the authors presented an alternate formalization and model checking approach for HyperMTL. These two works are quite similar in synchronous semantics; however, the formalization of asynchronous semantics was presented differently. While the asynchronous semantics in \cite{bonakdarpour2020model} is based on the the existence of an infinite sequence of timestamps and allows trace to proceed at different speeds, the asynchronous semantics of \cite{ho2018verifying} keeps a global clock in its analysis of traces and proceeds in order.  Model checking \cite{clarke1997model} has extensively been used to verify hyperproperties of models abstracted as transition systems by examining their related state transition graphs \cite{clarke2001bounded}. In \cite{finkbeiner2015algorithms}, the first model checking algorithms for HyperLTL and HyperCTL$^{*}$ employing alternating automata were proposed, which was also adopted in \cite{bonakdarpour2020model, ho2018verifying}  to verify HyperMTL properties. An extensive study on the complexity of verifying hyperproperties with model checking is presented in \cite{bonakdarpour2018complexity}. Our formulation of HyperTWTL is closely related to the HyperMTL \cite{bonakdarpour2020model} proposed to express timed hyperproperties in discrete-time systems. However, as previously mentioned in introduction, classical TWTL formalism has several advantages including compactness obtained through the use of concatenation operator ($\ser$) which is very useful in robotic applications. 
%
A security-aware robotic motion planning approach was recently proposed using synchronous HyperTWTL specification and SMT solvers~\cite{10137643}. In contrast to\cite{10137643}, this paper presents an adequate version of HyperTWTL and addresses the model checking problem for both synchronous and asynchronous HyperTWTL.
\section{Conclusion}
\label{conclusion}

This paper introduced a model checking algorithm for a hyper-temporal logic, HyperTWTL, with synchronous and asynchronous semantics. Using a Technical Surveillance Squadron (TESS) case study, we showed that HyperTWTL can express important properties related to information-flow security policies and concurrency in complex robotic systems. Our proposed model checking algorithm verifies fragments of HyperTWTL by reducing the problem to a TWTL model checking problem. In the future, we plan to propose methods and algorithms for monitoring and synthesizing the alternation-free and $k$-alternations fragments of HyperTWTL.

\bibliographystyle{ACM-Reference-Format}
\bibliography{Ref}


\begin{thebibliography}{49}


\ifx \showCODEN    \undefined \def \showCODEN     #1{\unskip}     \fi
\ifx \showDOI      \undefined \def \showDOI       #1{#1}\fi
\ifx \showISBNx    \undefined \def \showISBNx     #1{\unskip}     \fi
\ifx \showISBNxiii \undefined \def \showISBNxiii  #1{\unskip}     \fi
\ifx \showISSN     \undefined \def \showISSN      #1{\unskip}     \fi
\ifx \showLCCN     \undefined \def \showLCCN      #1{\unskip}     \fi
\ifx \shownote     \undefined \def \shownote      #1{#1}          \fi
\ifx \showarticletitle \undefined \def \showarticletitle #1{#1}   \fi
\ifx \showURL      \undefined \def \showURL       {\relax}        \fi
\providecommand\bibfield[2]{#2}
\providecommand\bibinfo[2]{#2}
\providecommand\natexlab[1]{#1}
\providecommand\showeprint[2][]{arXiv:#2}

\bibitem[Agrawal and Bonakdarpour(2016)]%
        {agrawal2016runtime}
\bibfield{author}{\bibinfo{person}{Shreya Agrawal} {and}
  \bibinfo{person}{Borzoo Bonakdarpour}.} \bibinfo{year}{2016}\natexlab{}.
\newblock \showarticletitle{Runtime verification of k-safety hyperproperties in
  HyperLTL}. In \bibinfo{booktitle}{\emph{2016 IEEE 29th Computer Security
  Foundations Symposium (CSF)}}. IEEE, \bibinfo{pages}{239--252}.
\newblock


\bibitem[Aksaray et~al\mbox{.}(2021)]%
        {aksaray2021probabilistically}
\bibfield{author}{\bibinfo{person}{Derya Aksaray}, \bibinfo{person}{Yasin
  Yaz{\i}c{\i}o{\u{g}}lu}, {and} \bibinfo{person}{Ahmet~Semi Asarkaya}.}
  \bibinfo{year}{2021}\natexlab{}.
\newblock \showarticletitle{Probabilistically guaranteed satisfaction of
  temporal logic constraints during reinforcement learning}. In
  \bibinfo{booktitle}{\emph{2021 IEEE/RSJ International Conference on
  Intelligent Robots and Systems (IROS)}}. IEEE, \bibinfo{pages}{6531--6537}.
\newblock


\bibitem[Alahmadi et~al\mbox{.}(2022)]%
        {alahmadi2022cyber}
\bibfield{author}{\bibinfo{person}{Adel~N Alahmadi}, \bibinfo{person}{Saeed~Ur
  Rehman}, \bibinfo{person}{Husain~S Alhazmi}, \bibinfo{person}{David~G Glynn},
  \bibinfo{person}{Hatoon Shoaib}, {and} \bibinfo{person}{Patrick Sol{\'e}}.}
  \bibinfo{year}{2022}\natexlab{}.
\newblock \showarticletitle{Cyber-Security Threats and Side-Channel Attacks for
  Digital Agriculture}.
\newblock \bibinfo{journal}{\emph{Sensors}} \bibinfo{volume}{22},
  \bibinfo{number}{9} (\bibinfo{year}{2022}), \bibinfo{pages}{3520}.
\newblock


\bibitem[Alpern and Schneider(1985)]%
        {alpern1985defining}
\bibfield{author}{\bibinfo{person}{Bowen Alpern} {and} \bibinfo{person}{Fred~B
  Schneider}.} \bibinfo{year}{1985}\natexlab{}.
\newblock \showarticletitle{Defining liveness}.
\newblock \bibinfo{journal}{\emph{Information processing letters}}
  \bibinfo{volume}{21}, \bibinfo{number}{4} (\bibinfo{year}{1985}),
  \bibinfo{pages}{181--185}.
\newblock


\bibitem[Asarkaya et~al\mbox{.}(2021a)]%
        {asarkaya2021persistent}
\bibfield{author}{\bibinfo{person}{Ahmet~Semi Asarkaya}, \bibinfo{person}{Derya
  Aksaray}, {and} \bibinfo{person}{Yasin Yazicioglu}.}
  \bibinfo{year}{2021}\natexlab{a}.
\newblock \showarticletitle{Persistent aerial monitoring under unknown
  stochastic dynamics in pick-up and delivery missions}. In
  \bibinfo{booktitle}{\emph{AIAA Scitech 2021 Forum}}. \bibinfo{pages}{1125}.
\newblock


\bibitem[Asarkaya et~al\mbox{.}(2021b)]%
        {asarkaya2021temporal}
\bibfield{author}{\bibinfo{person}{Ahmet~Semi Asarkaya}, \bibinfo{person}{Derya
  Aksaray}, {and} \bibinfo{person}{Yasin Yaz{\i}c{\i}o{\u{g}}lu}.}
  \bibinfo{year}{2021}\natexlab{b}.
\newblock \showarticletitle{Temporal-logic-constrained hybrid reinforcement
  learning to perform optimal aerial monitoring with delivery drones}. In
  \bibinfo{booktitle}{\emph{2021 International Conference on Unmanned Aircraft
  Systems (ICUAS)}}. IEEE, \bibinfo{pages}{285--294}.
\newblock


\bibitem[Barthe et~al\mbox{.}(2004)]%
        {self-composition}
\bibfield{author}{\bibinfo{person}{G. Barthe}, \bibinfo{person}{P.R.
  D'Argenio}, {and} \bibinfo{person}{T. Rezk}.}
  \bibinfo{year}{2004}\natexlab{}.
\newblock \showarticletitle{Secure information flow by self-composition}. In
  \bibinfo{booktitle}{\emph{Proceedings. 17th IEEE Computer Security
  Foundations Workshop, 2004.}} \bibinfo{pages}{100--114}.
\newblock
\urldef\tempurl%
\url{https://doi.org/10.1109/CSFW.2004.1310735}
\showDOI{\tempurl}


\bibitem[Bonakdarpour et~al\mbox{.}(2018a)]%
        {bonakdarpour2018opportunities}
\bibfield{author}{\bibinfo{person}{Borzoo Bonakdarpour},
  \bibinfo{person}{Jyotirmoy~V Deshmukh}, {and} \bibinfo{person}{Miroslav
  Pajic}.} \bibinfo{year}{2018}\natexlab{a}.
\newblock \showarticletitle{Opportunities and challenges in monitoring
  cyber-physical systems security}. In \bibinfo{booktitle}{\emph{International
  Symposium on Leveraging Applications of Formal Methods}}. Springer,
  \bibinfo{pages}{9--18}.
\newblock


\bibitem[Bonakdarpour and Finkbeiner(2018)]%
        {bonakdarpour2018complexity}
\bibfield{author}{\bibinfo{person}{Borzoo Bonakdarpour} {and}
  \bibinfo{person}{Bernd Finkbeiner}.} \bibinfo{year}{2018}\natexlab{}.
\newblock \showarticletitle{The complexity of monitoring hyperproperties}. In
  \bibinfo{booktitle}{\emph{2018 IEEE 31st Computer Security Foundations
  Symposium (CSF)}}. IEEE, \bibinfo{pages}{162--174}.
\newblock


\bibitem[Bonakdarpour and Finkbeiner(2020)]%
        {bonakdarpour2020controller}
\bibfield{author}{\bibinfo{person}{Borzoo Bonakdarpour} {and}
  \bibinfo{person}{Bernd Finkbeiner}.} \bibinfo{year}{2020}\natexlab{}.
\newblock \showarticletitle{Controller synthesis for hyperproperties}. In
  \bibinfo{booktitle}{\emph{2020 IEEE 33rd Computer Security Foundations
  Symposium (CSF)}}. IEEE, \bibinfo{pages}{366--379}.
\newblock


\bibitem[Bonakdarpour et~al\mbox{.}(2020)]%
        {bonakdarpour2020model}
\bibfield{author}{\bibinfo{person}{Borzoo Bonakdarpour},
  \bibinfo{person}{Pavithra Prabhakar}, {and} \bibinfo{person}{C{\'e}sar
  S{\'a}nchez}.} \bibinfo{year}{2020}\natexlab{}.
\newblock \showarticletitle{Model checking timed hyperproperties in
  discrete-time systems}. In \bibinfo{booktitle}{\emph{NASA Formal Methods
  Symposium}}. Springer, \bibinfo{pages}{311--328}.
\newblock


\bibitem[Bonakdarpour et~al\mbox{.}(2018b)]%
        {bonakdarpour2018monitoring}
\bibfield{author}{\bibinfo{person}{Borzoo Bonakdarpour}, \bibinfo{person}{Cesar
  Sanchez}, {and} \bibinfo{person}{Gerardo Schneider}.}
  \bibinfo{year}{2018}\natexlab{b}.
\newblock \showarticletitle{Monitoring hyperproperties by combining static
  analysis and runtime verification}. In
  \bibinfo{booktitle}{\emph{International Symposium on Leveraging Applications
  of Formal Methods}}. Springer, \bibinfo{pages}{8--27}.
\newblock


\bibitem[Bonnah and Hoque(2022)]%
        {bonnah2022runtime}
\bibfield{author}{\bibinfo{person}{Ernest Bonnah} {and}
  \bibinfo{person}{Khaza~Anuarul Hoque}.} \bibinfo{year}{2022}\natexlab{}.
\newblock \showarticletitle{Runtime Monitoring of Time Window Temporal Logic}.
\newblock \bibinfo{journal}{\emph{IEEE Robotics and Automation Letters}}
  \bibinfo{volume}{7}, \bibinfo{number}{3} (\bibinfo{year}{2022}),
  \bibinfo{pages}{5888--5895}.
\newblock


\bibitem[Bonnah et~al\mbox{.}(2023)]%
        {10137643}
\bibfield{author}{\bibinfo{person}{Ernest Bonnah}, \bibinfo{person}{Luan
  Nguyen}, {and} \bibinfo{person}{Khaza~Anuarul Hoque}.}
  \bibinfo{year}{2023}\natexlab{}.
\newblock \showarticletitle{Motion Planning Using Hyperproperties for Time
  Window Temporal Logic}.
\newblock \bibinfo{journal}{\emph{IEEE Robotics and Automation Letters}}
  \bibinfo{volume}{8}, \bibinfo{number}{8} (\bibinfo{year}{2023}),
  \bibinfo{pages}{4386--4393}.
\newblock
\urldef\tempurl%
\url{https://doi.org/10.1109/LRA.2023.3280830}
\showDOI{\tempurl}


\bibitem[Brett et~al\mbox{.}(2017)]%
        {brett2017rewriting}
\bibfield{author}{\bibinfo{person}{Noel Brett}, \bibinfo{person}{Umair
  Siddique}, {and} \bibinfo{person}{Borzoo Bonakdarpour}.}
  \bibinfo{year}{2017}\natexlab{}.
\newblock \showarticletitle{Rewriting-based runtime verification for
  alternation-free HyperLTL}. In \bibinfo{booktitle}{\emph{International
  Conference on Tools and Algorithms for the Construction and Analysis of
  Systems}}. Springer, \bibinfo{pages}{77--93}.
\newblock


\bibitem[B{\"u}y{\"u}kko{\c{c}}ak et~al\mbox{.}(2021)]%
        {buyukkoccak2021distributed}
\bibfield{author}{\bibinfo{person}{Ali~Tevfik B{\"u}y{\"u}kko{\c{c}}ak},
  \bibinfo{person}{Derya Aksaray}, {and} \bibinfo{person}{Yasin Yazicioglu}.}
  \bibinfo{year}{2021}\natexlab{}.
\newblock \showarticletitle{Distributed Planning of Multi-Agent Systems with
  Coupled Temporal Logic Specifications}. In \bibinfo{booktitle}{\emph{AIAA
  Scitech 2021 Forum}}. \bibinfo{pages}{1123}.
\newblock


\bibitem[Chowdhury et~al\mbox{.}(2017)]%
        {chowdhury2017survey}
\bibfield{author}{\bibinfo{person}{Abdullahi Chowdhury}, \bibinfo{person}{Gour
  Karmakar}, {and} \bibinfo{person}{Joarder Kamruzzaman}.}
  \bibinfo{year}{2017}\natexlab{}.
\newblock \showarticletitle{Survey of recent cyber security attacks on robotic
  systems and their mitigation approaches}.
\newblock In \bibinfo{booktitle}{\emph{Detecting and Mitigating Robotic Cyber
  Security Risks}}. \bibinfo{publisher}{IGI global}, \bibinfo{pages}{284--299}.
\newblock


\bibitem[Clarke et~al\mbox{.}(2001)]%
        {clarke2001bounded}
\bibfield{author}{\bibinfo{person}{Edmund Clarke}, \bibinfo{person}{Armin
  Biere}, \bibinfo{person}{Richard Raimi}, {and} \bibinfo{person}{Yunshan
  Zhu}.} \bibinfo{year}{2001}\natexlab{}.
\newblock \showarticletitle{Bounded model checking using satisfiability
  solving}.
\newblock \bibinfo{journal}{\emph{Formal methods in system design}}
  \bibinfo{volume}{19}, \bibinfo{number}{1} (\bibinfo{year}{2001}),
  \bibinfo{pages}{7--34}.
\newblock


\bibitem[Clarke(1997)]%
        {clarke1997model}
\bibfield{author}{\bibinfo{person}{Edmund~M Clarke}.}
  \bibinfo{year}{1997}\natexlab{}.
\newblock \showarticletitle{Model checking}. In
  \bibinfo{booktitle}{\emph{International Conference on Foundations of Software
  Technology and Theoretical Computer Science}}. Springer,
  \bibinfo{pages}{54--56}.
\newblock


\bibitem[Clarkson et~al\mbox{.}(2014)]%
        {clarkson2014temporal}
\bibfield{author}{\bibinfo{person}{Michael~R Clarkson}, \bibinfo{person}{Bernd
  Finkbeiner}, \bibinfo{person}{Masoud Koleini}, \bibinfo{person}{Kristopher~K
  Micinski}, \bibinfo{person}{Markus~N Rabe}, {and} \bibinfo{person}{C{\'e}sar
  S{\'a}nchez}.} \bibinfo{year}{2014}\natexlab{}.
\newblock \showarticletitle{Temporal logics for hyperproperties}. In
  \bibinfo{booktitle}{\emph{International Conference on Principles of Security
  and Trust}}. Springer, \bibinfo{pages}{265--284}.
\newblock


\bibitem[Clarkson and Schneider(2010)]%
        {clarkson2010hyperproperties}
\bibfield{author}{\bibinfo{person}{Michael~R Clarkson} {and}
  \bibinfo{person}{Fred~B Schneider}.} \bibinfo{year}{2010}\natexlab{}.
\newblock \showarticletitle{Hyperproperties}.
\newblock \bibinfo{journal}{\emph{Journal of Computer Security}}
  \bibinfo{volume}{18}, \bibinfo{number}{6} (\bibinfo{year}{2010}),
  \bibinfo{pages}{1157--1210}.
\newblock


\bibitem[Coenen et~al\mbox{.}(2019)]%
        {coenen2019verifying}
\bibfield{author}{\bibinfo{person}{Norine Coenen}, \bibinfo{person}{Bernd
  Finkbeiner}, \bibinfo{person}{C{\'e}sar S{\'a}nchez}, {and}
  \bibinfo{person}{Leander Tentrup}.} \bibinfo{year}{2019}\natexlab{}.
\newblock \showarticletitle{Verifying hyperliveness}. In
  \bibinfo{booktitle}{\emph{International Conference on Computer Aided
  Verification}}. Springer, \bibinfo{pages}{121--139}.
\newblock


\bibitem[Finkbeiner et~al\mbox{.}(2020)]%
        {finkbeiner2020synthesis}
\bibfield{author}{\bibinfo{person}{Bernd Finkbeiner},
  \bibinfo{person}{Christopher Hahn}, \bibinfo{person}{Philip Lukert},
  \bibinfo{person}{Marvin Stenger}, {and} \bibinfo{person}{Leander Tentrup}.}
  \bibinfo{year}{2020}\natexlab{}.
\newblock \showarticletitle{Synthesis from hyperproperties}.
\newblock \bibinfo{journal}{\emph{Acta informatica}} \bibinfo{volume}{57},
  \bibinfo{number}{1} (\bibinfo{year}{2020}), \bibinfo{pages}{137--163}.
\newblock


\bibitem[Finkbeiner et~al\mbox{.}(2017a)]%
        {finkbeiner2017eahyper}
\bibfield{author}{\bibinfo{person}{Bernd Finkbeiner},
  \bibinfo{person}{Christopher Hahn}, {and} \bibinfo{person}{Marvin Stenger}.}
  \bibinfo{year}{2017}\natexlab{a}.
\newblock \showarticletitle{EAHyper: satisfiability, implication, and
  equivalence checking of hyperproperties}. In
  \bibinfo{booktitle}{\emph{International Conference on Computer Aided
  Verification}}. Springer, \bibinfo{pages}{564--570}.
\newblock


\bibitem[Finkbeiner et~al\mbox{.}(2018)]%
        {finkbeiner2018model}
\bibfield{author}{\bibinfo{person}{Bernd Finkbeiner},
  \bibinfo{person}{Christopher Hahn}, {and} \bibinfo{person}{Hazem Torfah}.}
  \bibinfo{year}{2018}\natexlab{}.
\newblock \showarticletitle{Model checking quantitative hyperproperties}. In
  \bibinfo{booktitle}{\emph{International Conference on Computer Aided
  Verification}}. Springer, \bibinfo{pages}{144--163}.
\newblock


\bibitem[Finkbeiner et~al\mbox{.}(2017b)]%
        {finkbeiner2017verifying}
\bibfield{author}{\bibinfo{person}{Bernd Finkbeiner},
  \bibinfo{person}{Christian M{\"u}ller}, \bibinfo{person}{Helmut Seidl}, {and}
  \bibinfo{person}{Eugen Z{\u{a}}linescu}.} \bibinfo{year}{2017}\natexlab{b}.
\newblock \showarticletitle{Verifying security policies in multi-agent
  workflows with loops}. In \bibinfo{booktitle}{\emph{Proceedings of the 2017
  ACM SIGSAC Conference on Computer and Communications Security}}.
  \bibinfo{pages}{633--645}.
\newblock


\bibitem[Finkbeiner et~al\mbox{.}(2015)]%
        {finkbeiner2015algorithms}
\bibfield{author}{\bibinfo{person}{Bernd Finkbeiner}, \bibinfo{person}{Markus~N
  Rabe}, {and} \bibinfo{person}{C{\'e}sar S{\'a}nchez}.}
  \bibinfo{year}{2015}\natexlab{}.
\newblock \showarticletitle{Algorithms for model checking HyperLTL and
  HyperCTL}. In \bibinfo{booktitle}{\emph{International Conference on Computer
  Aided Verification}}. Springer, \bibinfo{pages}{30--48}.
\newblock


\bibitem[Garey(1979)]%
        {garey1979computers}
\bibfield{author}{\bibinfo{person}{Michael~R Garey}.}
  \bibinfo{year}{1979}\natexlab{}.
\newblock \showarticletitle{computers and intractqbility}.
\newblock \bibinfo{journal}{\emph{A Guide to the Theory of NP-Completeness}}
  (\bibinfo{year}{1979}).
\newblock


\bibitem[Goguen and Meseguer(1982)]%
        {goguen1982security}
\bibfield{author}{\bibinfo{person}{Joseph~A Goguen} {and}
  \bibinfo{person}{Jos{\'e} Meseguer}.} \bibinfo{year}{1982}\natexlab{}.
\newblock \showarticletitle{Security policies and security models}. In
  \bibinfo{booktitle}{\emph{1982 IEEE Symposium on Security and Privacy}}.
  IEEE, \bibinfo{pages}{11--11}.
\newblock


\bibitem[Ho et~al\mbox{.}(2018)]%
        {ho2018verifying}
\bibfield{author}{\bibinfo{person}{Hsi-Ming Ho}, \bibinfo{person}{Ruoyu Zhou},
  {and} \bibinfo{person}{Timothy~M Jones}.} \bibinfo{year}{2018}\natexlab{}.
\newblock \showarticletitle{On verifying timed hyperproperties}.
\newblock \bibinfo{journal}{\emph{arXiv preprint arXiv:1812.10005}}
  (\bibinfo{year}{2018}).
\newblock


\bibitem[Ho et~al\mbox{.}(2021)]%
        {ho2021timed}
\bibfield{author}{\bibinfo{person}{Hsi-Ming Ho}, \bibinfo{person}{Ruoyu Zhou},
  {and} \bibinfo{person}{Timothy~M Jones}.} \bibinfo{year}{2021}\natexlab{}.
\newblock \showarticletitle{Timed hyperproperties}.
\newblock \bibinfo{journal}{\emph{Information and Computation}}
  \bibinfo{volume}{280} (\bibinfo{year}{2021}), \bibinfo{pages}{104639}.
\newblock


\bibitem[Hsu et~al\mbox{.}(2021)]%
        {hsu2021bounded}
\bibfield{author}{\bibinfo{person}{Tzu-Han Hsu}, \bibinfo{person}{C{\'e}sar
  S{\'a}nchez}, {and} \bibinfo{person}{Borzoo Bonakdarpour}.}
  \bibinfo{year}{2021}\natexlab{}.
\newblock \showarticletitle{Bounded model checking for hyperproperties}. In
  \bibinfo{booktitle}{\emph{International Conference on Tools and Algorithms
  for the Construction and Analysis of Systems}}. Springer,
  \bibinfo{pages}{94--112}.
\newblock


\bibitem[Koymans(1990)]%
        {koymans1990specifying}
\bibfield{author}{\bibinfo{person}{Ron Koymans}.}
  \bibinfo{year}{1990}\natexlab{}.
\newblock \showarticletitle{Specifying real-time properties with metric
  temporal logic}.
\newblock \bibinfo{journal}{\emph{Real-time systems}} \bibinfo{volume}{2},
  \bibinfo{number}{4} (\bibinfo{year}{1990}), \bibinfo{pages}{255--299}.
\newblock


\bibitem[Lacava et~al\mbox{.}(2020)]%
        {lacava2020current}
\bibfield{author}{\bibinfo{person}{G Lacava}, \bibinfo{person}{A Marotta},
  \bibinfo{person}{F Martinelli}, \bibinfo{person}{A Saracino},
  \bibinfo{person}{A La~Marra}, \bibinfo{person}{E Gil-Uriarte}, {and}
  \bibinfo{person}{V~Mayoral Vilches}.} \bibinfo{year}{2020}\natexlab{}.
\newblock \showarticletitle{Current research issues on cyber security in
  robotics}.
\newblock  (\bibinfo{year}{2020}).
\newblock


\bibitem[Lera et~al\mbox{.}(2017)]%
        {lera2017cybersecurity}
\bibfield{author}{\bibinfo{person}{Francisco J~Rodr{\'\i}guez Lera},
  \bibinfo{person}{Camino~Fern{\'a}ndez Llamas},
  \bibinfo{person}{{\'A}ngel~Manuel Guerrero}, {and}
  \bibinfo{person}{Vicente~Matell{\'a}n Olivera}.}
  \bibinfo{year}{2017}\natexlab{}.
\newblock \showarticletitle{Cybersecurity of robotics and autonomous systems:
  Privacy and safety}.
\newblock \bibinfo{journal}{\emph{Robotics-legal, ethical and socioeconomic
  impacts}} (\bibinfo{year}{2017}).
\newblock


\bibitem[Luo et~al\mbox{.}(2020)]%
        {luo2020stealthy}
\bibfield{author}{\bibinfo{person}{Mulong Luo}, \bibinfo{person}{Andrew~C
  Myers}, {and} \bibinfo{person}{G~Edward Suh}.}
  \bibinfo{year}{2020}\natexlab{}.
\newblock \showarticletitle{Stealthy tracking of autonomous vehicles with cache
  side channels}. In \bibinfo{booktitle}{\emph{29th USENIX Security Symposium
  (USENIX Security 20)}}. \bibinfo{pages}{859--876}.
\newblock


\bibitem[Maler and Nickovic(2004)]%
        {maler2004monitoring}
\bibfield{author}{\bibinfo{person}{Oded Maler} {and} \bibinfo{person}{Dejan
  Nickovic}.} \bibinfo{year}{2004}\natexlab{}.
\newblock \showarticletitle{Monitoring temporal properties of continuous
  signals}.
\newblock In \bibinfo{booktitle}{\emph{Formal Techniques, Modelling and
  Analysis of Timed and Fault-Tolerant Systems}}.
  \bibinfo{publisher}{Springer}, \bibinfo{pages}{152--166}.
\newblock


\bibitem[Nguyen et~al\mbox{.}(2017)]%
        {nguyen2017hyperproperties}
\bibfield{author}{\bibinfo{person}{Luan~Viet Nguyen}, \bibinfo{person}{James
  Kapinski}, \bibinfo{person}{Xiaoqing Jin}, \bibinfo{person}{Jyotirmoy~V
  Deshmukh}, {and} \bibinfo{person}{Taylor~T Johnson}.}
  \bibinfo{year}{2017}\natexlab{}.
\newblock \showarticletitle{Hyperproperties of real-valued signals}. In
  \bibinfo{booktitle}{\emph{Proceedings of the 15th ACM-IEEE International
  Conference on Formal Methods and Models for System Design}}.
  \bibinfo{pages}{104--113}.
\newblock


\bibitem[Paviot-Adet et~al\mbox{.}(2022)]%
        {paviot2022structural}
\bibfield{author}{\bibinfo{person}{Emmanuel Paviot-Adet},
  \bibinfo{person}{Denis Poitrenaud}, \bibinfo{person}{Etienne Renault}, {and}
  \bibinfo{person}{Yann Thierry-Mieg}.} \bibinfo{year}{2022}\natexlab{}.
\newblock \showarticletitle{Structural Reductions and Stutter Sensitive
  Properties}.
\newblock \bibinfo{journal}{\emph{arXiv preprint arXiv:2212.04218}}
  (\bibinfo{year}{2022}).
\newblock


\bibitem[Peterson et~al\mbox{.}(2021)]%
        {peterson2021distributed}
\bibfield{author}{\bibinfo{person}{Ryan Peterson}, \bibinfo{person}{Ali~Tevfik
  Buyukkocak}, \bibinfo{person}{Derya Aksaray}, {and} \bibinfo{person}{Yasin
  Yaz{\i}c{\i}o{\u{g}}lu}.} \bibinfo{year}{2021}\natexlab{}.
\newblock \showarticletitle{Distributed safe planning for satisfying minimal
  temporal relaxations of TWTL specifications}.
\newblock \bibinfo{journal}{\emph{Robotics and Autonomous Systems}}
  \bibinfo{volume}{142} (\bibinfo{year}{2021}), \bibinfo{pages}{103801}.
\newblock


\bibitem[Pnueli(1977)]%
        {pnueli1977temporal}
\bibfield{author}{\bibinfo{person}{Amir Pnueli}.}
  \bibinfo{year}{1977}\natexlab{}.
\newblock \showarticletitle{The temporal logic of programs}. In
  \bibinfo{booktitle}{\emph{18th Annual Symposium on Foundations of Computer
  Science (sfcs 1977)}}. ieee, \bibinfo{pages}{46--57}.
\newblock


\bibitem[Romano({[n.\,d.]})]%
        {romano}
\bibfield{author}{\bibinfo{person}{Susan~A Romano}.}
  \bibinfo{year}{[n.\,d.]}\natexlab{}.
\newblock \bibinfo{title}{Persistent surveillance gives squadron its global
  purpose}.
\newblock
\newblock
\urldef\tempurl%
\url{https://www.af.mil/News/Article-Display/Article/1152329/persistent-surveillance-gives-squadron-its-global-purpose/}
\showURL{%
\tempurl}


\bibitem[Rusakov et~al\mbox{.}(2014)]%
        {rusakov2014simple}
\bibfield{author}{\bibinfo{person}{Andrey Rusakov}, \bibinfo{person}{Jiwon
  Shin}, {and} \bibinfo{person}{Bertrand Meyer}.}
  \bibinfo{year}{2014}\natexlab{}.
\newblock \showarticletitle{Simple concurrency for robotics with the Roboscoop
  framework}. In \bibinfo{booktitle}{\emph{2014 IEEE/RSJ International
  Conference on Intelligent Robots and Systems}}. IEEE,
  \bibinfo{pages}{1563--1569}.
\newblock


\bibitem[Simmons(1992)]%
        {simmons1992concurrent}
\bibfield{author}{\bibinfo{person}{Reid~G Simmons}.}
  \bibinfo{year}{1992}\natexlab{}.
\newblock \showarticletitle{Concurrent planning and execution for autonomous
  robots}.
\newblock \bibinfo{journal}{\emph{IEEE Control Systems Magazine}}
  \bibinfo{volume}{12}, \bibinfo{number}{1} (\bibinfo{year}{1992}),
  \bibinfo{pages}{46--50}.
\newblock


\bibitem[Stucki et~al\mbox{.}(2019)]%
        {stucki2019gray}
\bibfield{author}{\bibinfo{person}{Sandro Stucki}, \bibinfo{person}{C{\'e}sar
  S{\'a}nchez}, \bibinfo{person}{Gerardo Schneider}, {and}
  \bibinfo{person}{Borzoo Bonakdarpour}.} \bibinfo{year}{2019}\natexlab{}.
\newblock \showarticletitle{Gray-box monitoring of hyperproperties}. In
  \bibinfo{booktitle}{\emph{International Symposium on Formal Methods}}.
  Springer, \bibinfo{pages}{406--424}.
\newblock


\bibitem[Tkachev and Abate(2013)]%
        {tkachev2013formula}
\bibfield{author}{\bibinfo{person}{Ilya Tkachev} {and}
  \bibinfo{person}{Alessandro Abate}.} \bibinfo{year}{2013}\natexlab{}.
\newblock \showarticletitle{Formula-free finite abstractions for linear
  temporal verification of stochastic hybrid systems}. In
  \bibinfo{booktitle}{\emph{Proceedings of the 16th international conference on
  hybrid systems: Computation and control}}. \bibinfo{pages}{283--292}.
\newblock


\bibitem[Vasile et~al\mbox{.}(2016)]%
        {twtltool}
\bibfield{author}{\bibinfo{person}{C. Vasile}, \bibinfo{person}{D. Aksaray},
  {and} \bibinfo{person}{C. Belta}.} \bibinfo{year}{2016}\natexlab{}.
\newblock \bibinfo{title}{{PyTWTL} tool}.
\newblock
\newblock
\urldef\tempurl%
\url{https://sites.bu.edu/hyness/twtl/}
\showURL{%
Retrieved October 10, 2022 from \tempurl}


\bibitem[Vasile et~al\mbox{.}(2017)]%
        {vasile2017time}
\bibfield{author}{\bibinfo{person}{Cristian-Ioan Vasile},
  \bibinfo{person}{Derya Aksaray}, {and} \bibinfo{person}{Calin Belta}.}
  \bibinfo{year}{2017}\natexlab{}.
\newblock \showarticletitle{Time window temporal logic}.
\newblock \bibinfo{journal}{\emph{Theoretical Computer Science}}
  \bibinfo{volume}{691} (\bibinfo{year}{2017}), \bibinfo{pages}{27--54}.
\newblock


\bibitem[Zdancewic and Myers(2003)]%
        {zdancewic2003observational}
\bibfield{author}{\bibinfo{person}{Steve Zdancewic} {and}
  \bibinfo{person}{Andrew~C Myers}.} \bibinfo{year}{2003}\natexlab{}.
\newblock \showarticletitle{Observational determinism for concurrent program
  security}. In \bibinfo{booktitle}{\emph{16th IEEE Computer Security
  Foundations Workshop, 2003. Proceedings.}} IEEE, \bibinfo{pages}{29--43}.
\newblock


\end{thebibliography}

\end{document}